\documentclass[11pt]{article}
\usepackage[utf8]{inputenc}

\usepackage{parskip}
\usepackage[a4paper, total={6in, 9.5in}]{geometry}
\usepackage{graphicx}
\usepackage{amsmath, amsfonts}
\usepackage{amsthm}
\usepackage[bottom]{footmisc}
\usepackage{float}
\usepackage{enumitem}
\usepackage{bm}
\usepackage{array}
\usepackage{tikz}
\usepackage{xspace}

\newcommand{\bra}[1]{\langle#1|} 
\newcommand{\ket}[1]{|#1\rangle} 

\usetikzlibrary{matrix,positioning,chains,shapes.geometric, arrows, fit}
\tikzstyle{QKDmodule} = [rectangle, minimum width=1cm, minimum height=0.5cm, text centered, font=\scriptsize, draw=black]
\tikzstyle{KM} = [rectangle, minimum width=2cm, minimum height=0.5cm, text centered, font=\scriptsize, draw=black]
\tikzstyle{QKDN} = [rectangle, minimum width=10cm, minimum height=0.5cm, text centered, font=\scriptsize, draw=black]
\tikzstyle{arrowtwo} = [<->,>=stealth]
\tikzstyle{arrow} = [->,>=stealth]
\tikzstyle{node} = [rectangle, minimum width=2.2cm, minimum height=0.5cm, text centered, font=\scriptsize, draw=black!60, dashed]
\tikzstyle{app} = [rectangle, minimum width=2cm, minimum height=1cm, text centered, font=\scriptsize, draw=black]
\tikzstyle{desc} = [text centered, font=\scriptsize]
\tikzstyle{layer} = [rectangle ,minimum width=2cm, minimum height=0.5cm, align=right, font=\scriptsize]

\newfloat{experiment}{htbp}{loe}
\floatname{experiment}{Experiment}

\newcommand{\mathcmd}[1]{\ensuremath{#1}\xspace}

\newcounter{definition}
\renewcommand{\thedefinition}{\arabic{definition}}
\newenvironment{definition}{%
  \refstepcounter{definition}
  \par\medskip
  \noindent
  \textbf{Definition \thedefinition.}\quad
}{
  \par\medskip
}

\newcounter{theorem}
\renewcommand{\thetheorem}{\arabic{theorem}}
\newenvironment{theorem}{%
  \refstepcounter{theorem}
  \par\medskip
  \noindent
  \textbf{Theorem \thetheorem.}\quad
}{
  \par\medskip
}

\newcounter{lemma}
\renewcommand{\thelemma}{\arabic{lemma}}
\newenvironment{lemma}{%
  \refstepcounter{lemma}
  \par\medskip
  \noindent
  \textbf{Lemma \thelemma.}\quad
}{
  \par\medskip
}

\newcounter{corollary}
\renewcommand{\thecorollary}{\arabic{corollary}}

\usepackage[style=numeric-comp, sorting=none]{biblatex}
\title{Information-Theoretic Solutions for Seedless QRNG Bootstrapping and Hybrid PQC-QKD Key Combination \vspace{6mm}}

\author{Juan Antonio Vieira Giestinhas$^{1}$\footnote{Correspondence: qph504@york.ac.uk} , Timothy Spiller$^1$
\\[3mm]{\centering $^1$ University of York, YO10 5DD, York, United Kingdom}}
\date{}

\begin{document}

\maketitle

\begin{abstract}
This paper considers two challenges faced by practical quantum networks: the bootstrapping of seedless Quantum Random Number Generators (QRNGs) and the resilient combination of Post-Quantum Cryptography (PQC) and Quantum Key Distribution (QKD) keys. These issues are addressed using universal hash functions as strong seeded extractors, with security foundations provided by the Quantum Leftover Hash Lemma (QLHL). First, the 'randomness loop' in QRNGs -- the requirement of an initial random seed to generate further randomness -- is resolved by proposing a bootstrapping method using raw data from two independent sources of entropy, given by seedless QRNG sources. Second, it is argued that strong seeded extractors are an alternative to XOR-based key combining that presents different characteristics. Unlike XORing, our method ensures that if the combined output and one initial key are compromised, the remaining key material retains quantifiable smooth min-entropy and remains secure in exchange of longer keys. Furthermore, the proposed method allows binding of transcript information with key material in a natural way, providing a tool to replace computationally based combiners to extend ITS security of the initial key material to the final combined output. By modeling PQC keys as having HILL (Hastad, Impagliazzo, Levin and Luby) entropy, the framework is extended to hybrid PQC-QKD systems. This unified approach provides a mathematically rigorous and future-proof mechanism for both randomness generation and secure key management against quantum adversaries.
\end{abstract}

\section{Introduction}

The transition to quantum-safe communication infrastructure introduces new architectural requirements for high-assurance security. Central to these requirements are two distinct challenges: the 'bootstrapping' of initial randomness for hardware and the resilient combination of diverse cryptographic keys. While these problems appear unrelated, they both find an information-theoretic solution in the framework of strong seeded extractors~\cite{recent_devpment_extractors_2004}. Strong seeded extractors play a crucial role in modern cryptography to provide unconditionally secure randomness extractors~\cite{randomness_extract_modular_frame_Trevisan, trevisan_ext_portm_renn, General_LHL_Quantum_Tomamichel_2011, Postprocessing_QRNG_Toeplitz_Trevisan} - that is, given some classical input, with a characterized min-entropy, that has interacted with an unconditional adversary, and a uniform random (or weak~\cite{deficit_seed_randomness}) seed, the output of the randomness extractor is statistically close to uniform random conditioned to any adversary. The most well known strong seeded extractors proven to be unconditionally secure are the Trevisan's~\cite{trevisan_ext_portm_renn, randomness_extract_modular_frame_Trevisan} extractors and any universal hash function~\cite{universal-hash-functions} with the compression level given by the Quantum Leftover Hash Lemma (QLHL)~\cite{General_LHL_Quantum_Tomamichel_2011}, guaranteeing an unconditionally secure extraction. This work focuses on the modified Toeplitz family, which is proven to be universal$_2$ (and dual universal$_2$)~\cite{deficit_seed_randomness}, and thus a valid candidate to be a strong seeded extractor following the QLHL. This work addresses two challenges relevant for quantum networking.

{\bf Application 1}: Random number generators (RNGs) are essential components across various domains, ranging from gaming and gambling to cryptographic systems. Traditionally, RNGs relied on pseudo-random number generators (PRNGs), which are inherently incapable of providing information-theoretic security (ITS). In contrast, seeded Quantum Random Number Generators (QRNGs) can achieve unconditional security, provided that the post-processing phase employs a strong seeded extractor and the input min-entropy is characterized by a rigorous physical model~\cite{Postprocessing_QRNG_Toeplitz_Trevisan, QRNG_0}. While some QRNG architectures avoid post-processing entirely, for example in~\cite{QRNG_no_post_process}, many high-security applications require it, such as~\cite{Postprocessing_QRNG_Toeplitz_Trevisan, QRNG_0, QRNG_1, QRNG_2}. This work addresses the ``randomness loop", a scenario where two independent QRNGs lack an initial seed for post-processing, by proposing an ITS solution for seedless bootstrapping. In this framework, one QRNG provides the entropy input while the second provides an independent, albeit imperfect, seed. By applying the Quantum Leftover Hash Lemma (QLHL) with a weak seed~\cite{deficit_seed_randomness}, a bitstring close to uniform random can be extracted, provided the combined min-entropy of the input and seed is sufficiently high and no correlations exist between the sources. This application relates to the randomness extraction of two independent weak sources problem~\cite{two_source_quantum}.

{\bf Application 2}: A less common application of the randomness extractors given by the QLHL, not found in the literature, is the resilient combination of hybrid keys, such as mixing Post-Quantum Cryptography (PQC) and Quantum Key Distribution (QKD) keys, to provide unconditional security even if one source is compromised. A typical method to combine keys of the same length is by XORing them (\cite{nist_combination_keys}, section 6.3). While the XORing method is in theory one of the best methods to combine keys - being information-theoretic secure -, bad implementations could lead to double XORing keys, which corresponds to the no mixing case. Furthermore, the proposed method, that relies on the modified Toeplitz family and the QLHL, allows combination of the keys of arbitrary length such that if one of the keys and the hybrid key are revealed, the remaining key is still unconditionally secure following some arbitrary parameter - which is not possible if XOR is used. Compared to the XOR method, the disadvantage comes in the form of longer keys and a worse computational complexity that goes from $O(n)$, required for the XOR, to a matrix multiplication that has a complexity of $O(n^2)$ (or at best $O(n\log n)$ if the Fourier transformation is used for the Toeplitz family), where $n$ is the input length in bits. The QLHL is further modified to take into account the statistical closeness to uniform random (and smoothness parameter) of the seed. Other methods to combine key material include Key Derivation Functions (KDFs) or PRNGs~\cite{nist_combination_keys}, which are computationally secure. While these methods require less key material, they lack ITS; in contrast, our proposed method maintains ITS by accepting longer key lengths. 
While the presented tool can be used to combine keys as the XOR method does, an additional authentication mechanism is required to maintain the strongest security guarantees when combining different keys with different security levels, as discussed in~\cite{Hovelmanns_QKD_Oracle}. The additional authentication mechanism is not commented in this work and only the key combination part is discussed --- informally speaking, in the XOR-then-MAC approach discussed in~\cite{Hovelmanns_QKD_Oracle}, only the XOR part is commented, which can be replaced by the introduced tool of this work.

The applications 1 and 2 considered here deal with classical data, for which min-entropy can be quantified following some physical models. The QLHL along with the modified Toeplitz matrix results into a strong seeded extractor that by definition gives an uniform random output following some arbitrary security parameter, even against quantum adversaries, making the randomness extraction an unconditionally secure process despite the classical nature of the input (and the seed) - the QLHL takes into account the possibility of an unbounded adversary that is able to perform any quantum attack, including coherent attacks, to the classical parameters ``input" and ``seed". The information that the unbounded adversary is not able to gather in terms of bits from the input and seed are quantified into the corresponding conditional min-entropies that, as mentioned, are quantified following some physical model. A necessary but not sufficient condition for the QLHL work is to ensure that no correlations between the input and the seed conditioned to any unbounded adversary exist, since no secure randomness extractors can be built from correlated input and seed~\cite{SANTHA198675, Vadha_book_2012}.

This work starts by presenting in section~\ref{definitions} the QLHL with weak seed along with related definitions, such as (smooth) min-entropy and statistical distances, and lemmas used in this work. The QLHL is further expanded, in order to take into account the smoothness parameter on the seed. Next, relevant comments on the modified Toeplitz family are made in section~\ref{Toepltiz_case}. Then, section~\ref{section_one_source_extraction} presents the application of the QLHL using the modified Toeplitz family to extract a statistically close to uniform random seed using two independent seedless QRNGs. The utilization of the QLHL and modified Toeplitz family as a method of mixing secure keys with private seed is presented in section~\ref{section_private_seed_mix}, with discussion of the case where a statistically close to uniform random seed is public given in the subsequent section~\ref{section_public_seed_mix}. The presented tool is applied to extend the ITS security of an initial QKD key through a hybrid authenticated key exchange protocol (for instance the Muckle\# protocol~\cite{KEMuckle}), presented in appendix~\ref{AppendixA}; however no security proof is provided and the scheme in appendix~\ref{AppendixA} is only a proposition to be analyzed further in a future work. A conclusion is given in the last section~\ref{conclusion}.

\section{Preamble/Definitions}\label{definitions}

In this section, the mathematical preliminaries are presented. Unless otherwise stated, the definitions and notation of the smooth entropy framework for quantum information theory as detailed in Tomamichel (2016)~\cite{Tomamichel_2016_book}, and Vadhan (2012) for the theory of randomness extractors~\cite{Vadha_book_2012}, are adopted.


\begin{definition}\label{length_variable}
    (Length or size of a variable) Let $X$ be a random variable defined in $\{0,1\}^n$. The length (or size) of the variable $X$ in terms of bits is noted in this work as $|X| = n$.
\end{definition}

\begin{definition}\label{def_classical_min_entropy}
    (Classical min-entropy~\cite{Tomamichel_2016_book}) Let $X$ be a random variable defined in $\{0,1\}^n$. The min-entropy of $X$ conditioned on a classical adversary/event $E$ is:
\end{definition}

\begin{equation*}
    H_{\text{min}}(X|E) = \sum_{e\in E}P_E(e) \max_{x \in X} P_{X|E=e}(x)
\end{equation*}

where $P_E(e)$ corresponds to the probability of having $e \in E$ and $P_{X|E=e}(x)$ is the probability of $x \in X$ given $e\in E$. The classical min-entropy quantifies the amount of randomness left in the variable $X$ given the presence of a classical adversary/event $E$, where adversaries use always the optimal strategy to guess $X$ given the classical information $E$. The classical version of the min-entropy is useful to explain what the equation corresponds to physically in a cryptographic context. However, to prove unconditional security, a generalization of the min-entropy that takes into account quantum adversaries is required.

\begin{definition}\label{def_quantum_min_entropy}
    (Min-entropy~\cite{General_LHL_Quantum_Tomamichel_2011, Tomamichel_2016_book}) Let $X$ be a random variable defined in $\{0,1\}^n$ and $\rho_{XE} \in S_{\leq}(\mathcal{H}_{XE})$, where $S_{\leq}(\mathcal{H}_{XE})$ is the set of sub-normalized quantum states defined on the bipartite Hilbert space $\mathcal{H}_{XE}$, $\left\{ \rho_{XE} \in \mathcal{P}(\mathcal{H}_{XE}) : 0 <\text{Tr}\rho_{XE}\leq1 \right\}$, and $\mathcal{P}(\mathcal{H}_{XE})$ is the set of positive semi-definite operators on $\mathcal{H}_{XE}$. The min-entropy of $X$ conditioned on an unconditional adversary/event $E$ is defined as:
\end{definition}

\begin{equation*}
    H_{\text{min}}(X|E)_{\rho} = \max_{\sigma_E \in S_{=}(\mathcal{H}_E)} \sup\left\{\lambda \in \mathbb{R}:\rho_{XE} \leq 2^{-\lambda}\mathbb{I}_X \otimes \sigma_E\right\}
\end{equation*}

Where $\rho = \rho_{XE}$, $\mathcal{H}_E$ is the finite-dimensional Hilbert space in which $E$ is acting and $S_=(\mathcal{H}_E)$ is the set of normalized quantum states $\left\{ \sigma_E \in \mathcal{P}(\mathcal{H}_E) : \text{Tr}\sigma_E=1 \right\}$.

If not mentioned otherwise, the considered min-entropies in this work are taken with an unconditional adversary $E$.

\begin{definition}\label{source_def}
    (Source) Let $X$ be a random variable defined in $\{0,1\}^n$ with min-entropy $H_{\text{min}}(X)$. Then $X$ is a $(|X|, H_{\text{min}}(X))$-source. Additionally, if the source generates the outcome with an adversary/event $E$ present then $X$ becomes a $(|X|, H_{\text{min}}(X|E))$-source.
\end{definition}

\begin{definition}\label{dynamic_source_def}
    (Dynamic source) Let $X$ be a random variable defined in $\{0,1\}^*$ with min-entropy $H_{\text{min}}(X)$ and arbitrary length $*= |X|$. Then $X$ is a $(|X|, H_{\text{min}}(X))$-source. Additionally, if the source generates the outcome with an adversary/event $E$ present then $X$ becomes a $(|X|, H_{\text{min}}(X|E))$-source.
\end{definition}

All sources in this work are considered to be dynamic unless stated otherwise.

\begin{definition}\label{definition_smooth_min_entropy}
    (Smooth min-entropy) In the presence of an uncertainty of $X$ conditioned on $E$, $\varepsilon \geq 0$, the $\varepsilon$-smooth min-entropy of $X$ conditioned on $E$ is defined as:
\end{definition}

\begin{equation*}
    H^\varepsilon_{\min}(X|E)_{\rho} = \max_{\tilde{\rho}_{XE} \in \mathcal{B}^\varepsilon(\rho_{XE})} H_{\min}(X|E)_{\tilde{\rho}}
\end{equation*}

where $\mathcal{B}^\varepsilon(\rho_{XE})$ is the $\varepsilon$-ball of states close to $\rho_{XE} \in S_{\leq}(\mathcal{H})$, $\{ \tilde{\rho}_{XE} \in S_{\leq}(\mathcal{H}_{XE}): P(\rho_{XE}, \tilde{\rho}_{XE}) \leq \varepsilon \}$, and $P(\rho_{XE}, \tilde{\rho}_{XE})$ is the purified distance between states $\rho_{XE}$ and $\tilde{\rho}_{XE}$ introduced in definition~\ref{definition_purified_distance}.

Since the ball of states is defined using purified distances, in the following of his work, when treating bitstrings that are $\varepsilon$-close to uniform random, a penalty on their smoothing penalty scaling with the square root of $\varepsilon$ is found -- this is shown in Theorem~\ref{eq:secure_source_theorem}. Removal of the penalty is not trivial and would be beneficial for efficiency reasons; this remains an active area of research.

\begin{definition}\label{definition_purified_distance}
    (Purified distance~\cite{General_LHL_Quantum_Tomamichel_2011}) The purified distance between two states $\rho,\tau \in S_{\leq}(\mathcal{H})$ is defined as:
\end{definition}

\begin{equation*}
    P(\rho, \tau) = \sqrt{1-\bar{F}(\rho, \tau)^2}
\end{equation*}

where $\bar{F}(\rho, \tau)$ is the generalized fidelity between states $\rho,\tau \in S_{\leq}(\mathcal{H})$, equal to $\text{Tr}\left| \sqrt{\rho} \sqrt{\tau} \right| + \sqrt{(1 - \text{Tr}\rho)(1 - \text{Tr}\tau)}$, and $\text{Tr}A$ is defined as the trace of matrix $A$. 

Since this work aims to hybridize secure keys by mixing them, the notion of security has to be commented too: a random variable $X$ is $\varepsilon$-close to undistinguishable from a uniform random set of the same support length. Classically, this means that the statistical distance between $X$ conditioned on a classical adversary and the uniform random set $U_{|X|}$ (definition~\ref{definition_uniform_random}) introduced in definition~\ref{statsitical_distance_def_quantum}, is at most $\varepsilon$.

\begin{definition}\label{definition_uniform_random}
    The uniform random set of bit length $|X|$ is noted as $U_{|X|}$ and is defined as:
\end{definition}

\begin{equation*}
    U_{|X|} = \left\{ u \in \{0,1\}^{|X|}: P_{U_{|X|}}(u) = \frac{1}{|X|} \right\}
\end{equation*}

Where $P_{U_{|X|}}(u)$ denotes the probability of having the element $u$ in the set $U_{|X|}$.

\begin{definition}\label{statsitical_distance_def}
    (Statistical distance from uniform - classical) Let $X$ be a random variable conditioned on classical $E$. The classical statistical distance from uniform of $X$ conditioned on $E$ is:
\end{definition}

\begin{equation*}
    \frac{1}{2}\left\| P_{X|E} - U_{|X|} \right\|_1 = \frac{1}{2} \sum_e P_E(e)\sum_x \left| P_{X|E}(x|e) - P_{U_{|X|}}(x) \right|
\end{equation*}

Where $P_E(e)$, $P_{X|E}(x|E)$ and $P_{U_{|X|}}(x)$ correspond to the probabilities of having elements $e \in E$, $x\in X$ conditioned on $E$, and $x \in U_{|X|}$, respectively.

\begin{definition}\label{statsitical_distance_def_quantum}\label{def:trace_distance}
    (Statistical distance - quantum) The statistical distance is also the trace distance when it comes to quantum states, that is, given the quantum states $\rho, \tau \in S_{\leq}(\mathcal{H})$, the trace distance is:
\end{definition}

\begin{equation}
    D(\rho, \tau) = \frac{1}{2}\left\| \rho - \tau \right\|_1 = \frac{1}{2}\text{Tr}\left| \rho - \tau \right|
\end{equation}

Security notions are extended to include quantum adversaries by working with distances from uniform and quantum states, definition 3 from~\cite{General_LHL_Quantum_Tomamichel_2011}, definition~\ref{distance_from_uniform} in this work. 

\begin{definition}\label{distance_from_uniform}
    (Distance from uniform) Let $\rho_{XE} \in S_{\leq}(\mathcal{H}_{XE})$. The distance from uniform of $X$ conditioned on $E$ is:
\end{definition}

\begin{equation*}
    D_u(X|E)_{\rho} = \min_{\sigma_E \in \mathcal{P}(\mathcal{H}_E)\,:\,\text{Tr}\sigma_E = \text{Tr}\rho_E} \frac{1}{2} \left\| \rho_{XE} - \frac{\mathbb{I}_X}{\dim \mathcal{H}_X} \otimes \sigma_E \right\|_1
\end{equation*}

If $D_u(X|E) \leq \varepsilon \ll1$ then $X$ is quantum secure and $\varepsilon$-close to uniform random. 

\begin{theorem}\label{theorem_secure_source}
    (Secure source) Let $X$ be $\varepsilon$-close to uniform random in the presence of quantum side information $E$ and let $\rho_{XE} \in S_{\leq}(\mathcal{H}_{XE})$. Then $X$ is called a $|X|_\varepsilon$-secure source where the smoothed min-entropy of $X$ conditioned to the quantum side information $E$ is:
    
\end{theorem}




\begin{equation}\label{eq:secure_source_theorem}
    H_{\min}^{\sqrt{2\varepsilon}}(X|E)_\rho = |X|
\end{equation}

\begin{proof}

$X$ is $\varepsilon$-close to uniform random, thus by definition:

\begin{equation*}
    D_u(X|E) = \min_{\sigma_E \in \mathcal{P}(\mathcal{H}_E)\,:\,\text{Tr}\sigma_E = \text{Tr}\rho_E} \frac{1}{2} \left\| \rho_{XE} - \frac{\mathbb{I}_X}{\dim \mathcal{H}_X} \otimes \sigma_E \right\|_1 \leq \varepsilon
\end{equation*}

Let $\Delta\left(\rho_{XE}, \frac{\mathbb{I}_X}{\dim \mathcal{H}_X} \otimes \sigma_E \right)$ be the generalized trace distance defined as:

\begin{equation*}
    \Delta\left(\rho_{XE}, \frac{\mathbb{I}_X}{\dim \mathcal{H}_X} \otimes \sigma_E \right) = \frac{1}{2} \left\| \rho_{XE} -  \frac{\mathbb{I}_X}{\dim \mathcal{H}_X} \otimes \sigma_E \right\|_1 + \frac{1}{2} \left|\text{Tr}\left( \rho_{XE} - \frac{\mathbb{I}_X}{\dim \mathcal{H}_X} \otimes \sigma_E \right) \right|
\end{equation*}

Furthermore, let $\sigma_E$ be constrained to be in $\mathcal{P}(\mathcal{H}_E)$ and $\text{Tr}\sigma_E = \text{Tr}\rho_E$.

Define $\Delta_0$ as the minimum over the possible quantum states $\sigma_E$:

\begin{equation*}
    \Delta_0 = \min_{\sigma_E \in \mathcal{P}(\mathcal{H}_E)\,:\,\text{Tr}\sigma_E = \text{Tr}\rho_E} \frac{1}{2} \left\| \rho_{XE} -  \frac{\mathbb{I}_X}{\dim \mathcal{H}_X} \otimes \sigma_E \right\|_1 + \frac{1}{2} \left|\text{Tr}\left( \rho_{XE} - \frac{\mathbb{I}_X}{\dim \mathcal{H}_X} \otimes \sigma_E \right) \right|
\end{equation*}

Furthermore, since $\text{Tr}(\rho_E) = \text{Tr}(\sigma_{E})$ then:

\begin{equation*}
    \text{Tr}(\sigma_E) = \text{Tr}(\rho_E) = \text{Tr}(\text{Tr}_X(\rho_{XE})) = \text{Tr}(\rho_{XE})
\end{equation*}

Additionally, because the term $\frac{\mathbb{I}_X}{\dim \mathcal{H}_X}$ belongs in a normalized set, then:

\begin{eqnarray*}
    \left|\text{Tr}\left( \rho_{XE} - \frac{\mathbb{I}_X}{\dim \mathcal{H}_X} \otimes \sigma_E \right) \right| &=& \left|\text{Tr}\left( \rho_{XE} \right) - \text{Tr}\left(\frac{\mathbb{I}_X}{\dim \mathcal{H}_X} \otimes \sigma_E \right) \right|\\
    &=& \left| \text{Tr}(\rho_E) - \text{Tr}(\sigma_E) \right| = 0
\end{eqnarray*}

Implying that the distance from uniform $D_u(X|E)$ is exactly the minimum that the generalized trace distance can achieve under the constrained states $\sigma_E$:

\begin{equation*}
    \Delta_0 = D_u(X|E)
\end{equation*}

Moreover, Lemma 3.17 from citation~\cite{Tomamichel_2016_book}, derived from the Fuchs and van de Graaf inequalities~\cite{Fuchs_Graaf_inequalities}, states that given quantum states $\rho, \tau \in \mathcal{S_{\leq}(\mathcal{H})}$ then:

\begin{equation*}
    P(\rho,\tau) \leq \sqrt{2\Delta(\rho,\tau)-\Delta(\rho,\tau)^2} \leq \sqrt{2\Delta(\rho,\tau)}
\end{equation*}

Where $P(\rho, \tau)$ is the purified distance.

Inputting the relevant parameters for the problem at hand yields:

\begin{equation*}
    P\left(\rho_{XE}, \frac{\mathbb{I}_X}{\dim \mathcal{H}_X} \otimes \sigma_E\right) \leq \sqrt{2\Delta\left(\rho_{XE},\frac{\mathbb{I}_X}{\dim \mathcal{H}_X} \otimes \sigma_E\right)}
\end{equation*}

Where $\sigma_E$ still belongs to $\mathcal{P}(\mathcal{H})_E$ and $\text{Tr} \sigma_E = \text{Tr}\rho_E$.

Given that $\Delta_0$ is the minimum of all possible values of $\Delta$ given the possible quantum states $\sigma_E$, the following holds:

\begin{equation*}
    P\left(\rho_{XE}, \frac{\mathbb{I}_X}{\dim \mathcal{H}_X} \otimes \sigma_E\right) \leq \sqrt{2\Delta_0}
\end{equation*}

And since $\Delta_0 = D_u(X|E) \leq \varepsilon$, then:

\begin{equation*}
    P\left(\rho_{XE}, \frac{\mathbb{I}_X}{\dim \mathcal{H}_X} \otimes \sigma_E\right) \leq \sqrt{2\varepsilon}
\end{equation*}

Which implies that $\frac{\mathbb{I}_X}{\dim \mathcal{H}_X } \otimes \sigma_E  \in \mathcal{B}^{\sqrt{2\varepsilon}}(\rho_{XE})$.

Finally, the min-entropy of the idealized uniform random quantum state $\frac{\mathbb{I}_X}{\dim \mathcal{H}_X } \otimes \sigma_E$ is $\log(\dim \mathcal{H}_X)$, which is maximal for any quantum state belonging to the Hilbert space $\mathcal{H}_X$ -- thus, by the definition of the smooth-min entropy:

\begin{equation*}
    H_{\min}^{\sqrt{2\varepsilon}}(X|E)_\rho = H_{\min}(X|E)_{\frac{\mathbb{I}_X}{\dim \mathcal{H}_X} \otimes \sigma_E} = \log(\dim \mathcal{H}_X)
\end{equation*}

The proof ends by noting that $\dim \mathcal{H}_X = 2^{|X|}$, giving equation~\ref{eq:secure_source_theorem}.

\end{proof}

Additionally, let $X_1$ and $X_2$ be $|X_1|_{\varepsilon_1}$- and $|X_2|_{\varepsilon_2}$-secure sources, respectively. Then:

\begin{eqnarray*}
    &H^{\sqrt{2\varepsilon_1}}_{\text{min}}(X_1|X_2,E) = H^{\sqrt{2\varepsilon_1}}_{\text{min}}(X_1|E) \\
    &H^{\sqrt{2\varepsilon_2}}_{\text{min}}(X_2|X_1,E) = H^{\sqrt{2\varepsilon_2}}_{\text{min}}(X_2|E) \\
\end{eqnarray*}

When it comes to manipulating secure keys given by different cryptographic primitives such as QKD or PQC, the keys are always kept in classical registers. Thus, PQC and QKD key material is considered to come from classical secure sources - assuming that QKD was correctly implemented and PQC is secure against quantum adversaries. 




\begin{definition}\label{definition_concatenation_indpt_variables}
    (Concatenation of conditionally independent variables) Let $X_1$ and $X_2$ be two conditionally independent $(|X_1|, H_{\text{min}}(X_1|E))$- and $(|X_2|, H_{\text{min}}(X_2|E))$-sources given an adversary/event E and let $X$ be the source constructed from the concatenation of $X_1$ and $X_2$, then $X$ is a $(|X_1|+|X_2|, H_{\text{min}}(X_1,X_2|E))$-source where the following holds (given the conditional independence):
\end{definition}

\begin{equation*}
    H_{\text{min}}(X_1,X_2|E) = H_{\text{min}}(X_1|E) + H_{\text{min}}(X_2|E)
\end{equation*}

Additionally:

\begin{eqnarray*}
     H_{\text{min}}(X_1|X_2, E) =  H_{\text{min}}(X_1|E)\\
      H_{\text{min}}(X_2|X_1, E) =  H_{\text{min}}(X_2|E)
\end{eqnarray*}

\begin{theorem}\label{theorem_decomposition}
Let $X$ be a classical $|X|_\varepsilon$-secure source and be decomposed into $X_1$ and $X_2$ sources such that the elements of $X$ are the concatenation of the elements of $X_1$ and $X_2$. Then $X_1$ and $X_2$ correspond to a $|X_1|_\varepsilon$-secure and $|X_2|_\varepsilon$-secure sources, respectively, and both sources are independent variables conditioned on $E$, following the smoothing parameter $\varepsilon \geq 0$, where:
\end{theorem}

\begin{eqnarray*}
    &H_{\text{min}}^{\sqrt{2\varepsilon}}(X|E) = H_{\text{min}}^{\sqrt{2\varepsilon}}(X_1|E) + H_{\text{min}}^{\sqrt{2\varepsilon}}(X_2|E) = |X| \\
    &H_{\text{min}}^{\sqrt{2\varepsilon}}(X_1|E, X_2) = H_{\text{min}}^{\sqrt{2\varepsilon}}(X_1|E) = |X_1| \\
    &H_{\text{min}}^{\sqrt{2\varepsilon}}(X_2|E, X_1) = H_{\text{min}}^{\sqrt{2\varepsilon}}(X_2|E) = |X_2|
\end{eqnarray*}

\begin{proof}

Since $X_1$ is obtained by tracing out $X_2$ from the joint system $X$, we apply the Monotonicity of the Trace Distance under Partial Trace. Let $\rho_{X_1 E} = \text{Tr}_{X_2}(\rho_{XE})$ and $\frac{\mathbb{I}_{X_1}}{2^{|X_1|}} = \text{Tr}_{X_2}\left(\frac{\mathbb{I}_{X}}{2^{|X|}}\right)$. Let $\sigma_E$ be the quantum state that satisfies the $\varepsilon$-closeness from uniform random for the joint source $X$ (Definition~\ref{distance_from_uniform} and Theorem~\ref{theorem_secure_source}).

\begin{eqnarray*}
\frac{1}{2} \left\| \rho_{X_1 E} - \frac{\mathbb{I}_{X_1}}{2^{|X_1|}} \otimes \sigma_E \right\|_1 &=& \frac{1}{2} \left\| \text{Tr}_{X_2}\left(\rho_{XE}\right) - \text{Tr}_{X_2}\left(\frac{\mathbb{I}_{X}}{2^{|X|}} \otimes \sigma_E\right) \right\|_1 \\
&\leq& \frac{1}{2} \left\| \rho_{XE} - \frac{\mathbb{I}_{X}}{2^{|X|}} \otimes \sigma_E \right\|_1
\end{eqnarray*}

The last inequality holds for all considered quantum states, thus the following holds as well:

\begin{equation*}
\min_{\sigma_E \in \mathcal{P}(\mathcal{H}_E)\,:\,\text{Tr}\sigma_E = \text{Tr}\rho_E} \frac{1}{2} \left\| \rho_{X_1 E} - \frac{\mathbb{I}_{X_1}}{2^{|X_1|}} \otimes \sigma_E \right\|_1 \leq \min_{\sigma_E \in \mathcal{P}(\mathcal{H}_E)\,:\,\text{Tr}\sigma_E = \text{Tr}\rho_E} \frac{1}{2} \left\| \rho_{XE} - \frac{\mathbb{I}_{X}}{2^{|X|}} \otimes \sigma_E \right\|_1 \leq \varepsilon
\end{equation*}

The same reasoning applies when treating $\rho_{X_2 E}$.

Since $X_1$ and $X_2$ are $\varepsilon$-close to uniform random to $U_{|X_1|}$ and $U_{|X_2|}$ respectively, the following holds (Theorem~\ref{theorem_secure_source}):

\begin{eqnarray*}
    H_{\text{min}}^{\sqrt{2\varepsilon}}(X_1|E) = |X_1| \\
    H_{\text{min}}^{\sqrt{2\varepsilon}}(X_2|E) = |X_2|
\end{eqnarray*}

Furthermore, since $|X| = |X_1| + |X_2|$ the following holds:

\begin{equation}
    H_{\text{min}}^{\sqrt{2\varepsilon}}(X|E) = H_{\text{min}}^{\sqrt{2\varepsilon}}(X_1|E) + H_{\text{min}}^{\sqrt{2\varepsilon}}(X_2|E)
\end{equation}

The second part of the proof stats by introducing a quantum state $\sigma_E^*$ such that the following is minimized:

\begin{eqnarray*}
    \min_{\sigma_E \in \mathcal{P}(\mathcal{H}_E)\,:\,\text{Tr}\sigma_E = \text{Tr}\rho_E} \frac{1}{2} \left\| \rho_{X_1X_2E} -  \frac{\mathbb{I}_{X_1}}{\dim \mathcal{H}_{X_1}} \otimes \frac{\mathbb{I}_{X_2}}{\dim \mathcal{H}_{X_2}} \otimes \sigma_E \right\|_1 \\
    =  \frac{1}{2} \left\| \rho_{X_1X_2E} -  \frac{\mathbb{I}_{X_1}}{\dim \mathcal{H}_{X_1}} \otimes \frac{\mathbb{I}_{X_2}}{\dim \mathcal{H}_{X_2}} \otimes \sigma_E^* \right\|_1
\end{eqnarray*}

Furthermore, notice that:

\begin{equation*}
    \text{Tr}\left( \frac{\mathbb{I}_{X_2}}{\dim \mathcal{H}_{X_2}} \otimes \sigma_E^* \right) = \text{Tr}\left(\sigma_E^* \right) = \text{Tr}\left(\rho_E \right)
\end{equation*}

Additionally, since $\text{Tr}\left(\rho_E \right) = \text{Tr}\left(\rho_{X_2E} \right)$, then:

\begin{equation*}
    \text{Tr}\left( \frac{\mathbb{I}_{X_2}}{\dim \mathcal{H}_{X_2}} \otimes \sigma_E^* \right)  = \text{Tr}\left(\rho_{X_2E} \right)
\end{equation*}

Consequently, the quantum state $\frac{\mathbb{I}_{X_2}}{\dim \mathcal{H}_{X_2}} \otimes \sigma_E^*\in \mathcal{P}(\mathcal{H}_{X_2E})$ is a candidate to minimize trace distance from uniform with respect to $\rho_{X_1X_2E}$, yielding:

\begin{eqnarray*}
    \min_{\sigma_{X_2E} \in \mathcal{P}(\mathcal{H}_{X_2E})\,:\,\text{Tr}\sigma_{X_2E} = \text{Tr}\rho_{X_2E}} \frac{1}{2} \left\| \rho_{X_1X_2E} -  \frac{\mathbb{I}_{X_1}}{\dim \mathcal{H}_{X_1}} \otimes \sigma_{X_2E} \right\|_1 \\
    \leq  \frac{1}{2} \left\| \rho_{X_1X_2E} -  \frac{\mathbb{I}_{X_1}}{\dim \mathcal{H}_{X_1}} \otimes \frac{\mathbb{I}_{X_2}}{\dim \mathcal{H}_{X_2}} \otimes \sigma_E^* \right\|_1
\end{eqnarray*}

Finally, by noting that the right side of the inequality corresponds to the minimum of the trace distance from uniform with respect to $\rho_{XE} = \rho_{X_1X_2E}$:

\begin{eqnarray*}
    \min_{\sigma_{X_2E} \in \mathcal{P}(\mathcal{H}_{X_2E})\,:\,\text{Tr}\sigma_{X_2E} = \text{Tr}\rho_{X_2E}} \frac{1}{2} \left\| \rho_{X_1X_2E} -  \frac{\mathbb{I}_{X_1}}{\dim \mathcal{H}_{X_1}} \otimes \sigma_{X_2E} \right\|_1 \\
    \leq  \frac{1}{2} \left\| \rho_{XE} -  \frac{\mathbb{I}_{X_1}}{\dim \mathcal{H}_{X_1}} \otimes \frac{\mathbb{I}_{X_2}}{\dim \mathcal{H}_{X_2}} \otimes \sigma_E^* \right\|_1 \leq \varepsilon
\end{eqnarray*}

Where the last inequality holds since $X$ is $\varepsilon$-close to uniform random. Given Theorem~\ref{theorem_secure_source}:

\begin{equation*}
    H_{\min}^{\sqrt{2\varepsilon}}(X_1 | X_2E) = |X_1| = H_{\min}^{\sqrt{2\varepsilon}}(X_1 | E)
\end{equation*}

The same is found when analyzing the smooth min-entropy of $X_2$ given quantum side information $X_1E$, ending the proof.

\end{proof}

Consequently, any two elements coming from a decomposition of an $|X|_\varepsilon$-source can be treated as independent variables conditioned on $E$ given the smoothness parameter $\sqrt{2\varepsilon}$.

\begin{theorem}\label{theorem_concatenation}
    Let $X_1$ and $X_2$ be two $|X_1|_{\varepsilon_1}$- and $|X_2|_{\varepsilon_2}$-secure sources, respectively, that are independent from all possible quantum side information $E_1$ and $E_2$. Denote by $X$ the set given by the concatenation of any bitstrings from $X_1$ and $X_2$. Then, for any quantum side information $E$, $X$ corresponds to a classical $|X|_{\varepsilon_1+\varepsilon_2}$-secure source where:
\end{theorem}

\begin{equation}\label{concatenate_independent_theorem}
    H^{\sqrt{2\varepsilon_1}}_{\text{min}}(X_1|E) + H^{\sqrt{2\varepsilon_2}}_{\text{min}}(X_2|E) = H^{\sqrt{2(\varepsilon_1 + \varepsilon_2)}}_{\text{min}}(X|E)
\end{equation}

\begin{proof}

Without the smoothing parameters, the sources are perfectly secure, meaning that $X_1$ and $X_2$ correspond to the uniform random distribution of size $|X_1|$ and $|X_2|$, respectively. Thus, for all adversaries $E$, given definition~\ref{definition_concatenation_indpt_variables}:

\begin{eqnarray*}
    &H_{\text{min}}(X_1|E) = |X_1| \\
    &H_{\text{min}}(X_2|E) = |X_2| \\
    &H_{\text{min}}(X_1|E) + H_{\text{min}}(X_2|E) = H_{\text{min}}(X|E) = |X| \\
\end{eqnarray*}

Incorporating the smoothing parameters, $\varepsilon_1$ and $\varepsilon_2$ with respect to $X_1$ and $X_2$ respectively, adds the following constraints as stated by Theorem~\ref{theorem_secure_source}:

\begin{eqnarray*}
    &H^{\sqrt{2\varepsilon_1}}_{\text{min}}(X_1|E) = |X_1| \\
    &H^{\sqrt{2\varepsilon_2}}_{\text{min}}(X_2|E) = |X_2| \\
\end{eqnarray*}

Moreover, 

\begin{equation*}
    H^{\sqrt{2\varepsilon_1}}_{\text{min}}(X_1|E) + H^{\sqrt{2\varepsilon_2}}_{\text{min}}(X_2|E) = H^{\varepsilon^*}_{\text{min}}(X|E) = |X|
\end{equation*}

Where $\varepsilon^*$ is upper-bounded by $\sqrt{2(\varepsilon_1 + \varepsilon_2)}$ as proved below.

Since $X_1$ and $X_2$ are independently conditioned on the adversary, the joint state of the system is the tensor product of the individual states. Let the joint Hilbert space be $\mathcal{H}_{X_1 E_1} \otimes \mathcal{H}_{X_2 E_2}$. Let $\rho_{X_1E_1}$ and $\rho_{X_2E_2}$ be the states of the first and second sources respectively, where $\rho_{X_1E_1} \otimes \rho_{X_2E_2} = \rho_{XE}$ ($X$ is the result of concatenating $X_1$ and $X_2$). Let $\sigma_1 = \frac{\mathbb{I}_{X_1}}{2^{|X_1|}} \otimes \sigma_{E_1^*}$ and $\sigma_2 = \frac{\mathbb{I}_{X_2}}{2^{|X_2|}} \otimes \sigma_{E_2^*}$ be the respective ideal states such that $\|\rho_{X_1E} - \sigma_1\|_1$ and $\|\rho_{X_2E} - \sigma_2\|_1$ are minimized over the possible states $\sigma_{E_1^*} \in \mathcal{P}(\mathcal{H}_{E_1})$ and $\sigma_{E_2^*}\in \mathcal{P}(\mathcal{H}_{E_2})$, where $\text{Tr}(\rho_{E_1}) = \text{Tr}(\sigma_{E_1^*})$ and $\text{Tr}(\rho_{E_2}) = \text{Tr}(\sigma_{E_2^*})$. The trace distance of the joint system is bounded using the Triangle Inequality by introducing the intermediate state $\rho_{X_1 E_1} \otimes \sigma_2$:

\begin{eqnarray*}
\frac{1}{2} \| \rho_{X_1E_1} \otimes \rho_{X_2 E_2} - \sigma_1 \otimes \sigma_2 \|_1&\leq& \frac{1}{2} \left\| \rho_{X_1 E_1} \otimes \rho_{X_2 E_2} - \rho_{X_1 E_1} \otimes \sigma_2 \right\|_1 + \frac{1}{2} \left\| \rho_{X_1 E_1} \otimes \sigma_2 - \sigma_1 \otimes \sigma_2 \right\|_1 \\
&=& \frac{1}{2} \left\| \rho_{X_1 E_1} \otimes (\rho_{X_2 E_2} - \sigma_2) \right\|_1 + \frac{1}{2} \left\| (\rho_{X_1 E_1} - \sigma_1) \otimes \sigma_2 \right\|_1
\end{eqnarray*}

Using the property that the trace norm is multiplicative on tensor products ($\|A \otimes B\|_1 = \|A\|_1 \|B\|_1$) and that density matrices are from sub-normalized sets:

\begin{align*}
&= \frac{1}{2} \|\rho_{X_1 E_1}\|_1 \|\rho_{X_2 E_2} - \sigma_2\|_1 + \frac{1}{2} \|\rho_{X_1 E_1} - \sigma_1\|_1 \|\sigma_2\|_1 \\
&\leq \frac{1}{2} \|\rho_{X_2E} - \sigma_2\|_1 + \frac{1}{2} \|\rho_{X_1E} - \sigma_1\|_1 \\
&\leq \varepsilon_2 + \varepsilon_1
\end{align*}

Where the final inequality holds because $\sigma_{E_1^*}$ and $\sigma_{E_2^*}$ minimize the $\|\rho_{X_1E} - \sigma_1\|_1$ and $\|\rho_{X_2E} - \sigma_2\|_1$, respectively, and $X_1$ and $X_2$ are assumed to be $|X_1|_{\varepsilon_1}$- and $|X_2|_{\varepsilon_2}$-secure sources. Given Theorem~\ref{theorem_secure_source} and by noting that $\rho_{X_1E_1} \otimes \rho_{X_2E_2} = \rho_{XE}$:

\begin{equation*}
    H^{\sqrt{2(\varepsilon_1 + \varepsilon_2)}}_{\min}(X|E) = |X|
\end{equation*}

Additionally, since $|X| = |X_1| + |X_2|$, then:

\begin{eqnarray*}
    H^{\sqrt{2(\varepsilon_1 + \varepsilon_2)}}_{\min}(X|E) = |X| &=& |X_1| + |X_2| \\
    &=& H^{\sqrt{2\varepsilon_1}}_{\min}(X_1|E) + H^{\sqrt{2\varepsilon_2}}_{\min}(X_2|E)
\end{eqnarray*}

Furthermore, due to the composability of the trace distance, any future information gained by the adversary that is correlated only with their existing quantum side information $E$ (and not directly with $X$) cannot increase this guessing probability beyond the bound derived from the initial state indistinguishability. Thus, $X$ is a $|X|_{\varepsilon_1 + \varepsilon_2}$-secure source, giving theorem~\ref{theorem_concatenation}.

\end{proof}

In practice, independently secure sources can be achieved for technologies that are known to be composable secure. QKD is an example of composable secure technology~\cite{benor2004universalcomposablesecurityquantum} -- the concatenation of two keys that are $\varepsilon_1$- and $\varepsilon_2$-secure is a key that is $\varepsilon_1+\varepsilon_2$-secure. Similarly, if QRNGs can be shown to be composable secure, then the same argument can be made when concatenating secret bitstrings from different iterations.

Dynamic and secure sources behave similarly to block sources (definition~\ref{block_source_def}) when the smoothness parameter is taken into account, see 
theorems~\ref{theorem_decomposition} and~\ref{theorem_concatenation}.

\begin{definition}\label{block_source_def}
    (Block Source) Let $X$ be a $(|X|, H_{\text{min}}(X|E))$-source. Let $X_1$ and $X_2$ be the output of source $X$ at iterations $1$ and $2$, respectively. The source is a called a block source if conditional independence between the outputs holds given $E$, that is:
\end{definition}

\begin{eqnarray*}
    &H_{\text{min}}(X_1, X_2|E) = H_{\text{min}}(X_1|E) + H_{\text{min}}(X_2|E) \\
\end{eqnarray*}

Furthermore, conditional independence of the outputs implies:

\begin{eqnarray*}
    &H_{\text{min}}(X_1|X_2,E) = H_{\text{min}}(X_1|E) \\
    &H_{\text{min}}(X_2|X_1,E) = H_{\text{min}}(X_2|E) \\
\end{eqnarray*}

Given its composable security, QKD keys are inherently keys that originate from a secure source which can act similar to a block source -- the same could be said for QRNGs if composable security is verified. However, PQC keys are not provable in terms of their min-entropy against unconditional adversaries, but given some assumption a HILL (Hastad, Impagliazzo, Levin and Luby) entropy can be defined, so that PQC keys originate from a (computationally) secure source. This is discussed in section~\ref{subsection_PQC_as_source}.

\begin{definition}
    ($(k,\varepsilon)$-seeded extractor~\cite{Extractor_definition}) Ext: $\{0,1\}^n \times \{0,1\}^d \rightarrow \{0,1\}^m$ is a $(k, \varepsilon)$-seeded extractor if for every $(|X|, H_{\text{min}}(X))$-source $X$, where $H_{\text{min}}(X) \geq k$,
\end{definition}

\begin{equation*}
    \|\text{Ext}(X,Y) - U_m\|_1 \leq \varepsilon
\end{equation*}

where $Y$ is uniformly distributed over $\{0,1\}^d$ and independent of $X$, and $U_m$ is the uniform distribution set over $\{0,1\}^m$. 

\begin{definition}
    ($(k,\varepsilon)$-strong seeded extractor~\cite{Extractor_definition}) Let Ext: $\{0,1\}^n \times \{0,1\}^d \rightarrow \{0,1\}^m$ be a $(k, \varepsilon)$-strong seeded extractor, where $H_{\text{min}}(X) \geq k$. Ext is strong if the following holds: 
\end{definition}

\begin{equation*}
    \left\|\text{Ext}(X,Y)\circ Y - U_{m}\circ Y\right\|_1 \leq \varepsilon
\end{equation*}

where $\circ$ denotes string concatenation (often noted as $||$ in works of cryptographic nature).

\begin{lemma}\label{Lemma_Leaked_seed}
    (Strong seeded extractors with a constrained min-entropy on the seed~\cite{deficit_seed_randomness}, theorem 6) Let Ext: $\{0,1\}^n \times \{0,1\}^d \rightarrow \{0,1\}^m$ be a $(k, \varepsilon)$-strong seeded extractor. Let $X$ be a $(n,k)$-source and let $Y$ be a source on $\{0,1\}^d$ with  min-entropy $d-\lambda$ (conditioned to quantum side information $E$). Then:
\end{lemma}

\begin{equation}
    \|\text{Ext}(X,Y) - U_m\|_1 \leq 2^\lambda \varepsilon
\end{equation}

This work focuses on the Toeplitz matrix family $F$, which is known to be a family of strong seeded extractors~\cite{Postprocessing_QRNG_Toeplitz_Trevisan} given by the Leftover Hash Lemma (universal$_2$ case)~\cite{General_LHL_Quantum_Tomamichel_2011}:

\begin{equation}\label{LHL}
    |output| \leq H^{\tilde{\varepsilon}}_{\text{min}}(input|E)-2\log\left(\frac{1}{\varepsilon}\right) + 2
\end{equation}

where $\tilde{\varepsilon}$ corresponds to the smoothing parameter given the min-entropy $H^{\tilde{\varepsilon}}_{\text{min}}(input|E)$. If the output length verifies the inequation~\ref{LHL} then the output bitstring is $(\tilde{\varepsilon} + \varepsilon)$- close to uniform random. If no smoothing parameter is involved, $\tilde{\varepsilon} =0$:

\begin{equation}\label{LHL_no_smootinhg}
    |output| \leq H_{\text{min}}(input|E)-2\log\left(\frac{1}{\varepsilon}\right) + 2
\end{equation}

Where $|output|$ corresponds to the length of the output given by an uniform randomly chosen Toeplitz matrix from the family $F$, $input$ corresponds to a $(|input|, H_{\text{min}}(input|E))$-source, and $\varepsilon$ corresponds to the security parameter that states how close the output is close to a uniform and random bit string. If inequation~\ref{LHL_no_smootinhg} holds then the output bitstring is $\varepsilon$-close to uniform random.

The inequality given by the Leftover Hash Lemma, inequation~\ref{LHL} (and~\ref{LHL_no_smootinhg}), can be further updated when the seed presents a constrained min-entropy strictly less than its length by following the inequality from lemma~\ref{Lemma_Leaked_seed}. Given a desired $\varepsilon'$ security parameter and a seed of length $|seed|$ with min-entropy $H_{\text{min}}(seed)=|seed| - \lambda$, the following equality has to hold:

\begin{equation*}
    \varepsilon'=2^\lambda\varepsilon
\end{equation*}

In other words, to have the desired $\varepsilon'$ security parameter to hold, $\varepsilon$ has to be equal to $2^{-\lambda}\varepsilon'$. A sound replacement in the Leftover Hash Lemma inequality~\ref{LHL} can be made, yielding:

\begin{equation*}
    |output| \leq  H^{\tilde{\varepsilon}}_{\text{min}}(input|E)-2\log(\frac{1}{2^{-\lambda}\varepsilon'}) + 2
\end{equation*}

The following is obtained by arranging the terms and replacing $\lambda=|seed|-H_{\text{min}}(seed|E)$:

\begin{equation}\label{LHL_leakedseed0}
    |output| \leq H^{\tilde{\varepsilon}}_{\text{min}}(input|E) + 2(H_{\text{min}}(seed|E) - |seed|)  - 2\log\left(\frac{1}{\varepsilon'}\right) + 2
\end{equation}

This same bound is found in~\cite{deficit_seed_randomness} (theorem 6), and~\cite{yan2022secureboundanalysisquantum} (theorem 1). If inequation $\ref{LHL_leakedseed0}$ holds, then the output bitstring is at least $(\tilde{\varepsilon} + \varepsilon')$-close to uniform random.

However, an improved bound regarding the penalty term of the seed can be derived, as shown in~\cite{deficit_seed_randomness} (theorem 7). Additionally, this work takes into account the potential smoothness parameters regarding the seed.

\begin{lemma}\label{LHL_lemma_mine}
    (Quantum leftover hash lemma with conditional independently smoothed input and seed, for 1-almost universal$_2$ hash family, extension of~\cite{General_LHL_Quantum_Tomamichel_2011} and~\cite{deficit_seed_randomness}) Let the input and seed of a 1-almost universal$_2$ hash family be generated through sources that are conditionally independent on unconditional adversary $E$. Let $\tilde{\varepsilon}$ and $\varepsilon_s$ be the smoothing parameters of the input and seed, respectively. Consider the case that the input and output density matrices, denoted $\rho_{XE} \in \mathcal{S}_{\leq}(\mathcal{H}_{XE})$ and $\rho_{FZE}\in \mathcal{S}_{\leq}(\mathcal{H}_{FZE})$ respectively, where $X$, $F$, $Z$ and $E$ represent the variable defining the input, 1-almost universal$_2$, output and quantum side information. are classical-quantum states of the form:
\end{lemma}

    \begin{equation}\label{eq:rho_input}
        \rho_{XE} = \sum_x \ket{x}\bra{x}_X \otimes \rho_E^{[x]}
    \end{equation}

    where the probability of $x$ occurring is the trace of the sub-normalized state $\rho_E^{[x]}$ and $\rho_E = \sum_x \rho_E^{[x]}$, and

    \begin{equation}\label{eq:rho_output}
        \rho_{FZE} = \sum_f \sum_x p_f \ket{f}\bra{f}_F \otimes  \ket{z}\bra{z}_Z \otimes \rho_E^{[f,z]}
    \end{equation}

    where $z$ is a bitstring of bit-size $l$, $p_f = 1/\dim(\mathcal{H}_F)$ and $\rho_E := \sum_{x, f(x) = z} \rho_E^{[x]}$.
    
    In this case, then for any $\tilde{\varepsilon} \geq 0$ and $\varepsilon_s \geq 0$:

\begin{equation}\label{General_LHL}
    |output| \leq  H^{\tilde{\varepsilon}}_{\text{min}}(input|E)_\rho + H^{\varepsilon_s}_{\text{min}}(seed|E)_\rho - |seed|  - 2\log\left(\frac{1}{\varepsilon'}\right) + 2
\end{equation}

where the output is $(\tilde{\varepsilon} + \varepsilon_s + \varepsilon')$-close to uniform random.

\begin{proof}

The proof takes and adapts Tomamichel's proof for the quantum leftover hash lemma~\cite{General_LHL_Quantum_Tomamichel_2011} along with theorem $7$ from~\cite{yan2022secureboundanalysisquantum} - the proof remains at the quantum level.

Let $X$, $F$, $Z$ and $E$ be the input set, universal$_2$ hash family set (the seeds are elements of such set), output set (given by applying hash functions from $F$ to the input from $X$) and unconditional adversary respectively. The distance from uniform conditioned on any adversary $E$ averaged over the constrained choices of the hash function $f \in F$ (where seed $=f$) is:

\begin{equation*}
    \Delta = \sum_fp_f D_u(Z|E)_{\rho^{[f]}}
\end{equation*}

Where $p_f$ is the probability of having the particular hash function $f$ (particular seed $=f$), and $D_u(Z|E)_{\rho^{[f]}} = \min_{\sigma_E} \frac{1}{2}\left\| \rho_{ZE} - \frac{\mathbb{I}}{\text{dim }\mathcal{H_Z}} \otimes \sigma_E \right\|_1$ is the distance from uniform of $Z$ conditioned on $E$ (definition~\ref{distance_from_uniform}) given the density operator $\rho_{ZE}$ chosen on $f$.

$\Delta$ is upper bounded by (equation (19) of~\cite{General_LHL_Quantum_Tomamichel_2011}):

\begin{equation*}
    \Delta \leq \frac{1}{2} \sum_f p_f \left\| \rho^{[f]}_{ZE}  - \frac{\mathbb{I}}{\text{dim }\mathcal{H}_Z}\otimes  \sigma_E\right\|_1 = D_u(Z|FE)_\rho
\end{equation*}

Where $\sigma_E$ is chosen accordingly to satisfy the equality.

Theorem 7 of~\cite{deficit_seed_randomness} (for 1-almost universal$_2$ hash functions) allows bringing back of the randomness extraction problem with non-uniform random universal$_2$ hash functions to the uniform random case, by taking into account a penalty term $2^{\frac{|seed| - H_{\min}(seed)}{2}}$, where the seed is assumed to be uncorrelated with the input. This problem is equivalent to the randomness extraction of a partially revealed seed, where the remaining non-revealed bits of the seed have a maximum min-entropy, where the assumption that the seed is not correlated with the input of the randomness extraction function holds. When combining with the QLHL~\cite{General_LHL_Quantum_Tomamichel_2011}, this yields:

\begin{equation*}
    2D_u(Z|FE)_\rho \leq \sqrt{2^{l-H_{\text{min}}(X|E)_{\rho}+|seed|-H_{\text{min}}(seed)_{\rho}}}
\end{equation*}

The replacement of $H_{\text{min}}(seed)$ with $H_{\text{min}}(seed|E)$ in the proof of theorem $7$ of~\cite{deficit_seed_randomness} is sound in this work, given that the input and seed are generated through sources that are conditionally independent on unconditional adversary $E$, thus:

\begin{equation*}
    2D_u(Z|FE)_\rho \leq \sqrt{2^{l-H_{\text{min}}(X|E)_{\rho}+|seed|-H_{\text{min}}(seed|E)_{\rho}}}
\end{equation*}

Now take into account the smoothing parameters regarding the seed and input, the same as is done in~\cite{General_LHL_Quantum_Tomamichel_2011}, noting conditional independence between the input and seed given an unbounded adversary $E$. Let $\tilde{\rho}_{XE} \in \mathcal{B}^{\tilde{\varepsilon}}(\rho_{XE})$ and $\tilde{\rho}_{FE} \in \mathcal{B}^{\varepsilon_s}(\rho_{FE})$ be the classical-quantum states that optimize the smooth min-entropy $H_{\text{min}}(X|E)_{\tilde{\rho}} = H^{\tilde{\varepsilon}}_{\text{min}}(X|E)_{\rho}$ and $H_{\text{min}}(seed|E)_{\tilde{\rho}} = H^{\varepsilon_s}_{\text{min}}(seed|E)_{\rho}$. Define $\tilde{\rho}_{FZE} = (\mathcal{A} \otimes \mathcal{I}_E)(\tilde{\rho}_{FXE})$, where $\mathcal{A}$ corresponds to the randomness extraction process and is a trace-preserving completely positive map from $\mathcal{H}_{FX} \rightarrow \mathcal{H}_{FZ}$ that maps $\tilde{\rho}_{FXE} \rightarrow (\mathcal{A \otimes \mathcal{I}}_E)(\tilde{\rho}_{FXE})$. A difference between~\cite{General_LHL_Quantum_Tomamichel_2011} and here is that $\rho_{FXE}$ cannot be separated into $\rho_F \otimes \rho_{XE}$ since the seed also depends on $E$, altogether while keeping the conditional independence on $X$ given $E$. By noting that the strong extractor can only decrease the purified distances in the quantum setting, the following holds:

\begin{equation*}
    \frac{1}{2}\left\| \rho_{FZE} - \tilde{\rho}_{FZE} \right\|_1 \leq P(\rho_{FZE}, \tilde{\rho}_{FZE}) \leq P(\rho_{FXE}, \tilde{\rho}_{FXE})
\end{equation*}

Since the seed and input are independently conditioned on unbounded $E$, the joint distance is bounded by the sum of the marginal distances:

\begin{equation*}
    \frac{1}{2}\left\| \rho_{FZE} - \tilde{\rho}_{FZE} \right\|_1 \leq P(\rho_{FXE}, \tilde{\rho}_{FXE}) \leq P(\rho_{FE}, \tilde{\rho}_{FE}) + P(\rho_{XE}, \tilde{\rho}_{XE}) \leq \varepsilon_s + \tilde{\varepsilon}
\end{equation*}

Furthermore, let $\tilde{\sigma}_{FE}$ be the state that minimizes the distance from uniform $D_u(Z|FE)_{\tilde{\rho}}$:

\begin{eqnarray*}
    2D_u(Z|FE)_{\rho} &\leq& \left\| \rho_{FZE} - \frac{\mathbb{I}}{\text{dim }\mathcal{H}}_Z \otimes \tilde{\sigma}_{FE} \right\|_1 \\
    &\leq& \left\| \rho_{FZE} - \tilde{\rho}_{FZE} \right\|_1 + \left\| \tilde{\rho}_{FZE} - \frac{\mathbb{I}}{\text{dim }\mathcal{H}}_Z \otimes \tilde{\sigma}_{FE}\right\|_1 \\
    &\leq& 2(\tilde{\varepsilon} + \varepsilon_s) + 2D_u(Z|FE)_{\tilde{\rho}} \\
    &\leq& 2(\tilde{\varepsilon} + \varepsilon_s) + \sqrt{2^{l-H_{\text{min}}(X|E)_{\tilde{\rho}}+|seed|-H_{\text{min}}(seed|E)_{\tilde{\rho}}}}
\end{eqnarray*}

Since $H_{\text{min}}(X|E)_{\tilde{\rho}} = H^{\tilde{\varepsilon}}_{\text{min}}(X|E)_{\rho}$ and $H_{\text{min}}(seed|E)_{\tilde{\rho}} = H^{\varepsilon_s}_{\text{min}}(seed|E)_{\rho}$:

\begin{eqnarray*}
    D_u(Z|FE)_{\rho} &\leq& \tilde{\varepsilon} + \varepsilon_s + \frac{1}{2} \sqrt{2^{l-H_{\text{min}}(X|E)_{\tilde{\rho}}+|seed|-H_{\text{min}}(seed|E)_{\tilde{\rho}}}} \\
    &\leq& \tilde{\varepsilon} + \varepsilon_s + \frac{1}{2} \sqrt{2^{l-H^{\tilde{\varepsilon}}_{\text{min}}(X|E)_{\rho}+|seed|-H^{\varepsilon_s}_{\text{min}}(seed|E)_{\rho}}}
\end{eqnarray*}

Which gives lemma~\ref{LHL_lemma_mine}, where $l=|output|$.

\end{proof}

Unless specified otherwise, the input and output density matrices involved with randomness extraction using the QLHL satisfy Equations~\ref{eq:rho_input} and~\ref{eq:rho_output} for the remainder of this work.

\section{Toeplitz Family case}\label{Toepltiz_case}

This works focus on the Toeplitz families as the universal$_2$ hash function to be used along with the QLHL, given their simplicity and ease to implement practically. 

\subsection{Typical Toeplitz Family}

When using the Toeplitz family the following constraints regarding the seed and input sizes enter into play:

\begin{equation*}
    |seed| = |output| + |input| - 1
\end{equation*}

Let $|X| = |seed| + |input|$, which yields:

\begin{eqnarray*}
    |seed| &=& \frac{|X|+|output|-1}{2}\\
    |input| &=& \frac{|X|-|output|+1}{2}
\end{eqnarray*}

Since $|output| \geq 1$ (otherwise there is no Toeplitz matrix), the size of the seed is always bigger than the input length in bits - for the Toeplitz matrix case.

\subsection{Modified Toeplitz Family}

This work uses the lemma~\ref{LHL_lemma_mine} to cover applications where uniform random bitstrings can be extracted in an unconditional manner, given some min-entropy assumptions. The modified Toeplitz matrix family from Appendix B of~\cite{deficit_seed_randomness} is a fitting candidate for application of lemma~\ref{LHL_lemma_mine}, since it is proven to be a $1$-almost universal$_2$ (and $1$-almost dual universal$_2$) family (\cite{deficit_seed_randomness}, Lemma $14$). The modified Toeplitz constraints are updated to be:

\begin{equation}\label{Toeplitz_Constr0}
    |seed| = |input| - 1
\end{equation}

Let $|X| = |seed| + |input|$, which yields:

\begin{eqnarray}\label{Toepltiz_Constr1}
    |seed| &=& \frac{|X|-1}{2}\\
    |input| &=& \frac{|X|+1}{2}\label{Toepltiz_Constr2}
\end{eqnarray}

The advantage of the modified Toeplitz matrix over the regular one is that the seed requires a reduced number of bits. 

\section{Application 1: extracting a secure seed from two conditionally independent QRNGs}\label{section_one_source_extraction}

Santha and Vazirani~\cite{SANTHA198675} show that extracting randomness out of a weak source alone is not secure, even if the source output is decomposed into ``input" and ``seed" and used with a strong seeded extractor. The intuition behind this is that correlations between the ``input" and the ``seed" could lead to the worst case scenario where the adversary learns the output, without really knowing the ``input" or the ``seed". Furthermore, the quantum leftover hash lemma also assumes that the ``input" and ``seed" have no correlations (so are conditionally independent); thus, the QLHL cannot be used when a singular output of a weak source is considered. However, the story is different if block sources (definition~\ref{block_source_def}) are considered.

Randomness extraction with a private seed using a singular source is possible with the following condition. Let $X$ be a dynamic (definition~\ref{dynamic_source_def}) and block (definition~\ref{block_source_def}) $(|X|, H_{\text{min}}(X|E))$-source. Let $X_1$ and $X_2$ be the output from $X$ on iterations $1$ and $2$. Since $X$ is a block source, $X_1$ and $X_2$ can be treated as conditionally independent sources. Then, an $\varepsilon'$-close uniform random bitstring can be obtained if the inequation given in theorem~\ref{LHL_lemma_mine} holds with $\tilde{\varepsilon}=\varepsilon_s=0$. Without loss of generality, let $X_1$ and $X_2$ be the $input$ and $seed$ respectively - the length of the bitstrings is given by the modified Toeplitz family constraints~\ref{Toepltiz_Constr1} and~\ref{Toepltiz_Constr2}.

The extraction of randomness using only a source is a well known problem in the randomness extraction literature and often leads to two-source randomness extraction, where the input and seed are generated by independent means~\cite{two_randomnd_extrac_ref1}.

Practically, technologies that allow users to have two independently sources are provably in a composable security framework: QKD links with public seed (for Privacy Amplification), or QRNG with pre-shared private seed are case examples. Furthermore, independent sources can be achieved by relaxing adversarial interactions: additional assumptions such as bounded adversaries, for example quantum memory-less adversaries~\cite{Damgrd2005CryptographyIT}, or computationally bounded adversaries, allow users to effectively have independent sources. 

A realistic scenario is to have two seedless QRNGs that extract randomness from two independent physical processes - but require a seed to post-process the raw bits into secure bits. Note that it is okay for the QRNGs to be the same version of the device, as long as their sources of randomness are independently. In this case, one of the seedless QRNGs would be used to generate the seed and the other seedless QRNG would be used to generate the input. Since the entropy source of both seedless QRNGs are assumed to be independently, where the unbounded adversary has no means to inject correlations, the seed and input are conditionally independent on unbounded adversary $E$. Let $X_1$ and $X_2$ be the bitstrings generated by the different independent and seedless QRNGs, with the respective min-entropies $H_{\text{min}}(X_1|E)$ and $H_{\text{min}}(X_2|E)$. Without loss of generality, let $X_1$ be the element used as input and $X_2$ as the seed. By taking the inequation~\ref{General_LHL} and manipulating the terms, the following condition must hold to have a positive output length $A \geq 0$, where $A$ is a natural number: 

\begin{eqnarray*}
    H_{\text{min}}(X_1|E) + H_{\text{min}}(X_2|E) \geq A + |X_2| + 2\log\left(\frac{1}{\varepsilon'}\right) - 2
\end{eqnarray*}

where the output is $\varepsilon'$-close uniform random. Additionally, the Toeplitz constraints~\ref{Toeplitz_Constr0} must hold if the dual universal hash family used is a Toeplitz family.

Note: If the bitstrings generated by the QRNGs are too long to be the seed or input, a truncation can be taken. Let $X^{trunc}$ be the truncation of $X$, where $|X^{trunc}| = n - q$ and $|X| = n$, then:

\begin{equation*}
    H_{\text{min}}(X|E) - q \leq H_{\text{min}}(X^{trunc}|E) \leq H_{\text{min}}(X|E)
\end{equation*}

Since no more information on the truncation and the leakage conditioned on $E$ is given, the worst case scenario has to be taken, that is:

\begin{equation*}
    H_{\text{min}}(X|E) - q = H_{\text{min}}(X^{trunc}|E)
\end{equation*}

\begin{figure}[ht]
    \centering
    \includegraphics[width=1\linewidth]{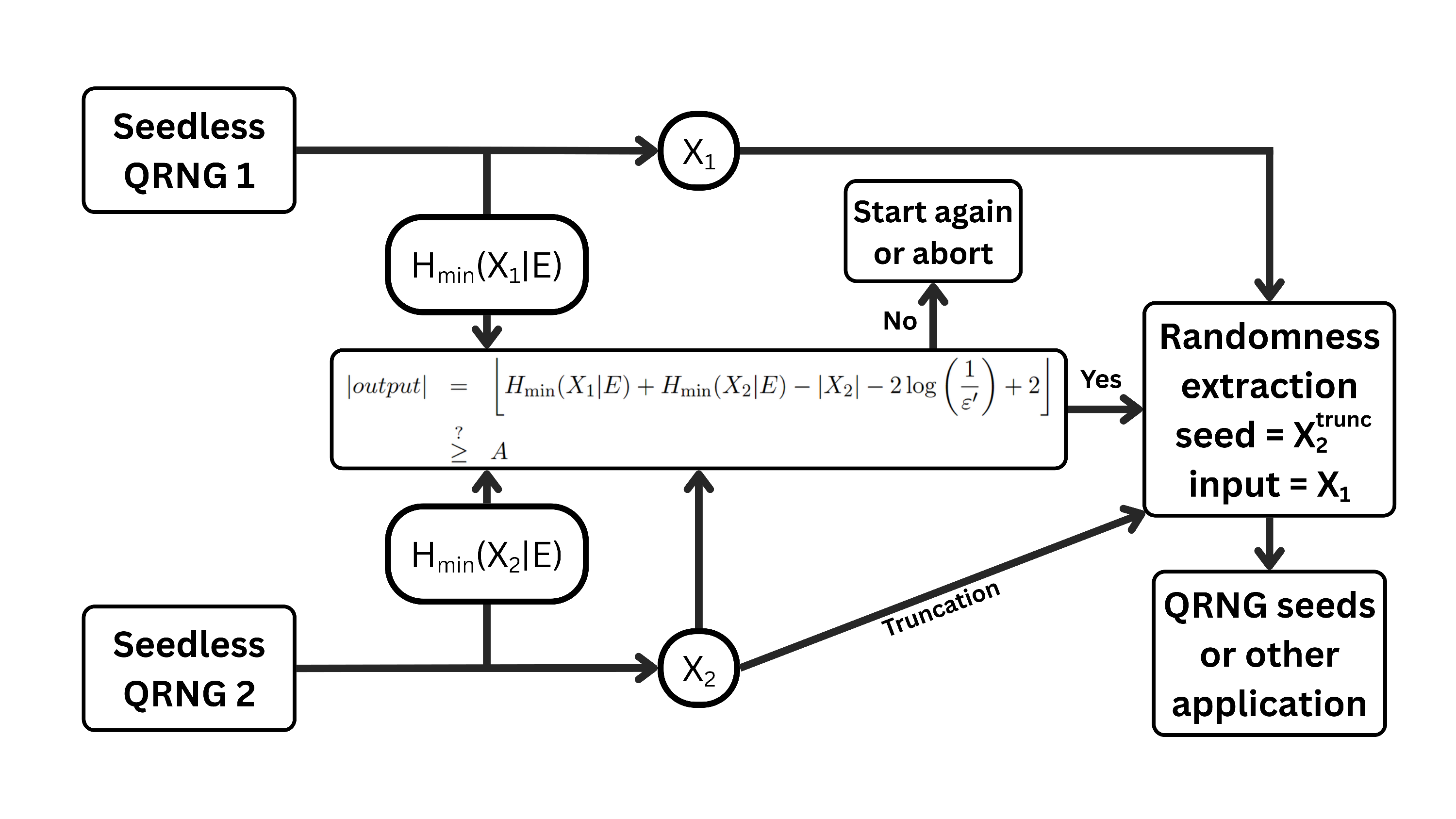}
    \caption{Overview of a strategy to extract a secure key under two seedless QRNGs.}
    \label{fig:seedless_QRNG}
\end{figure}

Summarized in Figure~\ref{fig:seedless_QRNG}, a strategy to apply a Toeplitz matrix where $X_2$ is the seed and $X_1$ the input is the following:

\begin{enumerate}
    \item Generate the bitstring $X_1$ (input) using one of the independent QRNGs, where $H_{\text{min}}(X_1|E)$ can be bounded following some appropriate physical model.
    \item Generate the bitstring $X_2$ (seed) using the other independent QRNG, where $H_{\text{min}}(X_2|E)$ can be bounded following some appropriate physical model and $|X_2| \geq C|X_1|$ where $C$ is a constant of the order of $O(1)$.
    \item Truncate $X_2$ $n$ times such that $|X_2| - n = |X_1| - 1$ to satisfy the Toeplitz constraint~\ref{Toeplitz_Constr0}.
    \item Verify that:
    \begin{equation*}
        H_{\text{min}}(X_1|E) + H_{\text{min}}(X_2|E) - n \geq A + |X_2| - n + 2\log\left(\frac{1}{\varepsilon'}\right) - 2
    \end{equation*}
    Since the number of truncations $n$ can be simplified on both sides of the inequation, it is sufficient to verify the inequality with the initial min-entropy of $X_2$ conditioned on $E$, $H_{\text{min}}(X_2|E)$, and its bit length $|X_2|$ (rather than the truncated terms). If the inequality holds, then the output has a length $|output|=A$ and is $\varepsilon'$-close to uniform random. Else, it is not possible to extract randomness given the min-entropies and steps 1 to 4 must be repeated with the hope of obtaining higher min-entropies for the $X_1$ and $X_2$ bitstrings.
\end{enumerate}

Application 2 is presented next, where the randomness extraction is used to combine two independent secure sources under private or public seed.

\section{Application 2: combining two independent secure sources - QKD and PQC keys case}\label{section_private_seed_mix}

The typical approach to use a strong seeded extractor is to send a uniform random seed through an authenticated channel. This case is discussed in section~\ref{section_public_seed_mix} after examining the two-source extraction case using the Generalized Quantum Leftover Hash Lemma~\ref{LHL_lemma_mine}, given that the input and seed are independently conditioned on any adversary $E$, where the seed used in the strong seeded extractor (Toeplitz family case) is constructed using the different sources, and hence private. Since the considered outputs of the sources are already secure according to some negligible smoothing parameters (epsilons), the randomness extraction of two secure sources corresponds to actually mixing secure keys.

\subsection{Two-secure source randomness extraction: combining keys}

Let $X_1$ and $X_2$ be two independent $|X_1|_{\varepsilon_1}$-secure and $|X_2|_{\varepsilon_2}$-secure sources respectively. Let us denote the source outputs as ``keys". The goal is to combine the output of both sources and examine the following cases:

\begin{enumerate}[label=\roman*)]
    \item If an adversary has control of the generation of one of the keys, then can the combined key be secure?\label{label1}
    \item If one of the keys is revealed after combining the keys, then can the combined key be secure?\label{label2}
    \item If the combined key is revealed, what is the remaining security level regarding the other two keys?\label{label3}
    \item If the combined key and one of the keys are revealed, what is the remaining security level regarding the other key?\label{label4}
\end{enumerate}

The case scenario where partial information on either keys is revealed is not treated in this work and can be easily obtained by replacing the corresponding min-entropies with leakage inside the presented formulas. 

In order to mitigate as much as possible the damage caused by revealing any of the key material, let both the seed and the input of the Toeplitz extractor be composed of the same $\alpha$ portion in terms of bit length of $X_1$ and $X_2$ - that is:

\begin{eqnarray}\label{intermediary_toeplt_1}
    &|seed|= \alpha(|X_1| + |X_2|) \\
    \label{intermediary_toeplt_2}&|input| = (1 - \alpha)(|X_1| + |X_2|)
\end{eqnarray}

Let $|X|=|seed|+|input|=|X_1| + |X_2|$ and by noting the modified Toeplitz constraints~\ref{Toepltiz_Constr1} and~\ref{Toepltiz_Constr2}:

\begin{eqnarray*}
    |seed| &=& \frac{|X_1|+|X_2|-1}{2}\\
    |input| &=& \frac{|X_1|+|X_2|+1}{2}
\end{eqnarray*}

Thus:

\begin{equation}\label{alpha_eq0}
    \alpha = \frac{|X_1|+|X_2| - 1}{2(|X_1|+|X_2|)}
\end{equation}

Since the keys $X_1$ and $X_2$ come from secure sources, the $seed$ and $input$ bitstrings have the following property given theorems~\ref{theorem_decomposition} and~\ref{theorem_concatenation}:

\begin{eqnarray*}
    &H^{\sqrt{2(\varepsilon_1 + \varepsilon_2)}}_{\text{min}}(seed|E) = H^{\sqrt{2\varepsilon_1}}_{\text{min}}(X^\alpha_1|E) + H^{\sqrt{2\varepsilon_2}}_{\text{min}}(X^\alpha_2|E) = \alpha(|X_1| + |X_2|) \\
    &H^{\sqrt{2(\varepsilon_1 + \varepsilon_2)}}_{\text{min}}(input|E) = H^{\sqrt{2\varepsilon_1}}_{\text{min}}(X^{(1-\alpha)}_1|E) + H^{\sqrt{2\varepsilon_2}}_{\text{min}}(X^{(1-\alpha)}_2|E) = (1-\alpha)(|X_1| + |X_2|)
\end{eqnarray*}

where $X^\alpha_1$ corresponds to part of the decomposed set from $X_1$ with $\alpha$ corresponding to the portion taken in number of bits. Given the equations~\ref{intermediary_toeplt_1} and~\ref{intermediary_toeplt_2}:

\begin{eqnarray*}
&H^{\sqrt{2(\varepsilon_1 + \varepsilon_2)}}_{\text{min}}(seed|E) = |seed| \\
    &H^{\sqrt{2(\varepsilon_1 + \varepsilon_2)}}_{\text{min}}(input|E) = |input|
\end{eqnarray*}

Furthermore, Theorem~\ref{theorem_concatenation} states that the $seed$ and $input$ are classical $|seed|_{\varepsilon_1 + \varepsilon_2}$- and $|input|_{\varepsilon_1 + \varepsilon_2}$-secure sources, respectively. Additionally, since the $\alpha$ and $(1-\alpha)$ portions of the $seed$ and $input$ can be considered independently conditioned on any $E$, and $X_1$ and $X_2$ are by definition independently conditioned on $E$, then the $seed$ and $input$ bitstrings can be considered as independently conditioned on any $E$ while taking into account the smoothness parameters. Given theorem~\ref{theorem_decomposition} on the $seed$ and $input$ sources:

\begin{eqnarray*}
    &H^{\sqrt{2(\varepsilon_1 + \varepsilon_2)}}_{\text{min}}(seed, input|E)=H^{\sqrt{2(\varepsilon_1 + \varepsilon_2)}}_{\text{min}}(seed|E) + H^{\sqrt{2(\varepsilon_1 + \varepsilon_2)}}_{\text{min}}(input|E) \\
    &H^{\sqrt{2(\varepsilon_1 + \varepsilon_2)}}_{\text{min}}(seed|E, input)=H^{\sqrt{2(\varepsilon_1 + \varepsilon_2)}}_{\text{min}}(seed|E) \\
    &H^{\sqrt{2(\varepsilon_1 + \varepsilon_2)}}_{\text{min}}(input|E, seed)= H^{\sqrt{2(\varepsilon_1 + \varepsilon_2)}}_{\text{min}}(input|E)
\end{eqnarray*}

which directly relates to the independence between the $seed$ and the $input$ secure sources conditioned on $E$. Thus, the generalized quantum leftover hash lemma giving lemma~\ref{LHL_lemma_mine} can be used to combine the keys.

\begin{figure}[ht]
    \centering
    \includegraphics[width=1\linewidth]{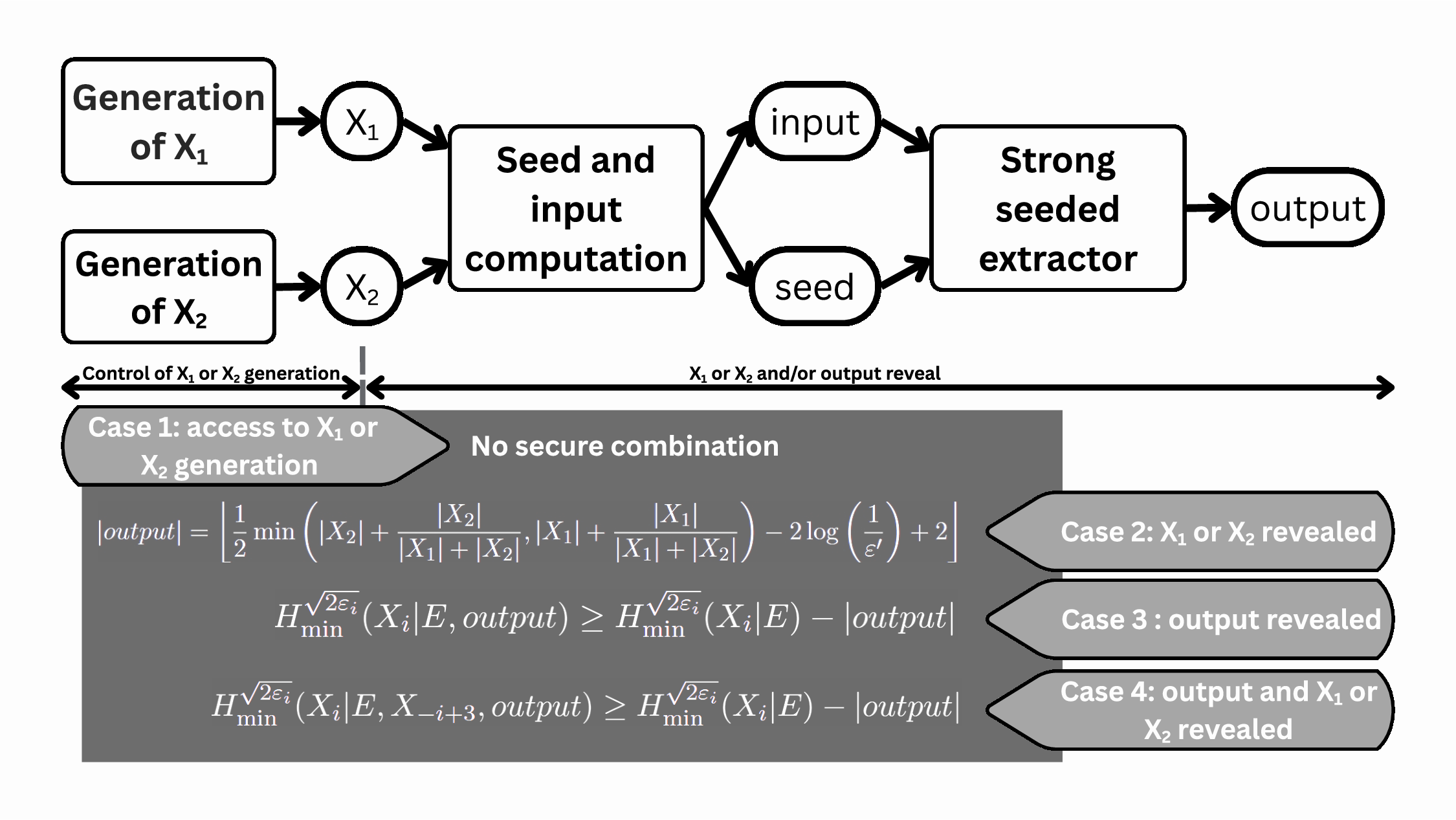}
    \caption{(Top) Overview of the two-secure source randomness extraction steps; (Bottom) Summary of the results for the different cases. $X_1$ and $X_2$ are considered to be $\varepsilon_1$- and $\varepsilon_2$-secure. The output is $(2{\sqrt{2(\varepsilon_1 + \varepsilon_2)}} + \varepsilon')$-close to uniform random.}
    \label{fig:two-secure-extraction}
\end{figure}

The top of figure~\ref{fig:two-secure-extraction} illustrates an overview of the steps to follow to execute the two-secure source randomness extraction. The bottom part of the figure summarizes the results obtained for each case.

For the omitted case where no key material nor output is compromised/revealed, replace all the terms in the inequation~\ref{General_LHL} to obtain the output bit length:

\begin{eqnarray*}
    |output| &\leq&  H^{\sqrt{2(\varepsilon_1 + \varepsilon_2)}}_{\text{min}}(input|E) + H^{\sqrt{2(\varepsilon_1 + \varepsilon_2)}}_{\text{min}}(seed|E) - |seed|  - 2\log\left(\frac{1}{\varepsilon'}\right) + 2 \\
    &=& |input| + |seed| - |seed|  - 2\log\left(\frac{1}{\varepsilon'}\right) + 2\\
    &=& |input| - 2\log\left(\frac{1}{\varepsilon'}\right) + 2\\
\end{eqnarray*}

where the output is at worst $(2{\sqrt{2(\varepsilon_1 + \varepsilon_2)}} + \varepsilon')$-close to uniform random.

Replacing $|input|$ given by the modified Toeplitz constraint~\ref{Toepltiz_Constr2} yields:

\begin{eqnarray*}
    |output| &\leq&  \frac{|X_1| + |X_2|+1}{2} - 2\log\left(\frac{1}{\varepsilon'}\right) + 2\\
\end{eqnarray*}

Thus:

\begin{eqnarray*}\label{output_floor0}
    |output| &=&  \left\lfloor\frac{|X_1| + |X_2|+1}{2} - 2\log\left(\frac{1}{\varepsilon'}\right) + 2 \right\rfloor \\
\end{eqnarray*}

\subsection{Case~\ref{label1}: Adversary has control of the generation of one of the keys}

Giving control of a key generation to an adversary is translated in this work as a decrease of the smooth min-entropy of the same key. Conditional independence on $E$ of the non-revealed information of the input and seed still holds, given that the two keys that compose the input and seed are independent. Thus, the Quantum Leftover Hash Lemma (QLHL) can still be formally applied to this conditional state to evaluate the remaining security bounds.

Without loss of generality, let $X_1$ be the key for which the adversary has control over its generation and let $X_2$ be secret - by symmetry, the same applies in the case the generation of $X_2$ is controlled and $X_1$ is secret. The smoothness of the min-entropies of the input and seed change as follows: 

\begin{eqnarray*}
    &H^{\sqrt{2(\varepsilon_1 + \varepsilon_2)}}_{\text{min}}(seed|E) =  H^{\sqrt{2\varepsilon_2}}_{\text{min}}(X^\alpha_2|E) = \alpha|X_2| \\
    &H^{\sqrt{2(\varepsilon_1 + \varepsilon_2)}}_{\text{min}}(input|E) = H^{\sqrt{2\varepsilon_2}}_{\text{min}}(X^{(1-\alpha)}_2|E) = (1-\alpha)|X_2|
\end{eqnarray*}

Replacing all the terms in inequation~\ref{General_LHL} yields:

\begin{eqnarray*}
    |output| &\leq&  H^{\sqrt{2(\varepsilon_1 + \varepsilon_2)}}_{\text{min}}(input|E) + H^{\sqrt{2(\varepsilon_1 + \varepsilon_2)}}_{\text{min}}(seed|E) - |seed|  - 2\log\left(\frac{1}{\varepsilon'}\right) + 2 \\
    &=& (1-\alpha)|X_2| + \alpha|X_2| - |seed|  - 2\log\left(\frac{1}{\varepsilon'}\right) + 2\\
\end{eqnarray*}

Replacing $|seed|$ given by the Toeplitz constraint~\ref{Toepltiz_Constr1} and arranging the terms yields:

\begin{eqnarray*}
    |output| &\leq& \frac{|X_2|-|X_1|+1}{2}  - 2\log\left(\frac{1}{\varepsilon'}\right) + 2\\
\end{eqnarray*}

Symmetrically, if the key $X_2$ generation is controlled by the adversary instead of the key $X_1$:

\begin{eqnarray*}
    |output| &\leq& \frac{|X_1|-|X_2|+1}{2}  - 2\log\left(\frac{1}{\varepsilon'}\right) + 2\\
\end{eqnarray*}

Since control by adversary of either of the keys is allowed, to have a positive output length, the following has to hold:

\begin{eqnarray*}
    0 &\leq& \frac{|X_2|-|X_1|+1}{2}  - 2\log\left(\frac{1}{\varepsilon'}\right) + 2\\
    0 &\leq& \frac{|X_1|-|X_2|+1}{2}  - 2\log\left(\frac{1}{\varepsilon'}\right) + 2\\
\end{eqnarray*}

And by summing both inequalities:

\begin{eqnarray*}
    0 &\leq& 5  - 4\log\left(\frac{1}{\varepsilon'}\right)\\
\end{eqnarray*} 

The right side of the inequation is negative when $\varepsilon' < 2^{-\frac{5}{4}}$, which typically is true. Hence, if either of the key generations is allowed to be controlled by an adversary, then no secure extraction (combining) is guaranteed.

In the context that one of the keys is allowed to be revealed but not the other, then a secure mixing could be achieved by changing how the seed and input are constructed, based on $X_1$, $X_2$ and which of the keys is allowed to be compromised.

Regarding this private seed scenario, a different strong extractor with seed length shorter than that required for Toeplitz families, such as the Trevisan's Extractors~\cite{Postprocessing_QRNG_Toeplitz_Trevisan}\cite{trevisan_ext_portm_renn}, would not really have the potential to surpass such inconvenience, since constructions for the weak seed case in reference~\cite{trevisan_ext_portm_renn} (section 5.4) also present a linear loss in the min-entropy of the seed. 

\subsection{Case~\ref{label2}: One of the keys is revealed after key generation}

Compared to the case~\ref{label1}, the smoothed min-entropy of the seed is fully kept in the equations: the adversary had no power to control over the key generation. The strong seeded extractor allows making partial, or full, reveal of the seed once randomness extraction has already happened, without affecting the security of the output. Naturally, the considered input corresponds to the portion given by the non-revealed key, which is independent from the seed.

Without loss of generality, let $X_1$ be the key that has been revealed and let $X_2$ be secret - by symmetry, the same applies in the case $X_2$ is revealed and $X_1$ is secret. The smoothness of the min-entropies of the input and seed change as follows: 

\begin{eqnarray*}
    &H^{\sqrt{2(\varepsilon_1 + \varepsilon_2)}}_{\text{min}}(seed|E) = H^{\sqrt{2\varepsilon_1}}_{\text{min}}(X^\alpha_1|E) + H^{\sqrt{2\varepsilon_2}}_{\text{min}}(X^\alpha_2|E) = \alpha(|X_1| + |X_2|) \\
    &H^{\sqrt{2(\varepsilon_1 + \varepsilon_2)}}_{\text{min}}(input|E) = H^{\sqrt{2\varepsilon_2}}_{\text{min}}(X^{(1-\alpha)}_2|E) = (1-\alpha)|X_2|
\end{eqnarray*}

Replacing all the terms in the inequation~\ref{General_LHL}:

\begin{eqnarray*}
    |output| &\leq&  H^{\sqrt{2(\varepsilon_1 + \varepsilon_2)}}_{\text{min}}(input|E) + H^{\sqrt{2(\varepsilon_1 + \varepsilon_2)}}_{\text{min}}(seed|E) - |seed|  - 2\log\left(\frac{1}{\varepsilon'}\right) + 2 \\
    &=& (1-\alpha)|X_2| + |seed| - |seed|  - 2\log\left(\frac{1}{\varepsilon'}\right) + 2\\
    &=& (1-\alpha)|X_2| - 2\log\left(\frac{1}{\varepsilon'}\right) + 2\\
\end{eqnarray*}

Replacing $\alpha$ (equation~\ref{alpha_eq0}) and rearranging the terms yields:

\begin{eqnarray*}
    |output| &\leq& \frac{|X_2|}{2} + \frac{|X_2|}{2(|X_1| + |X_2|)} - 2\log\left(\frac{1}{\varepsilon'}\right) + 2\\
\end{eqnarray*}

Symmetrically, if the key $X_2$ is revealed:

\begin{eqnarray*}
    |output| &\leq& \frac{|X_1|}{2} + \frac{|X_1|}{2(|X_1| + |X_2|)} - 2\log\left(\frac{1}{\varepsilon'}\right) + 2\\
\end{eqnarray*}

Giving a maximum output length of:

\begin{equation*}
    |output| = \left\lfloor \frac{1}{2}\min\left( |X_2| + \frac{|X_2|}{|X_1| + |X_2|}, |X_1| + \frac{|X_1|}{|X_1| + |X_2|} \right)  - 2\log\left(\frac{1}{\varepsilon'}\right) + 2 \right\rfloor
\end{equation*}

The output is $(2{\sqrt{2(\varepsilon_1 + \varepsilon_2)}} + \varepsilon')$-close to uniform random. 


\subsection{Case~\ref{label3}: Reveal output but no key}\label{only_output_reveal}

Since the parameter $output$ is classical, the following holds (\cite{Tomamichel_2016_book}, Lemma 6.18.):

\begin{eqnarray*}
    &H^{\sqrt{2\varepsilon_1}}_{\text{min}}(X_1|output, E) \geq H^{\sqrt{2\varepsilon_1}}_{\text{min}}(X_1|E) - |output| \\
    &H^{\sqrt{2\varepsilon_2}}_{\text{min}}(X_2|output, E) \geq H^{\sqrt{2\varepsilon_2}}_{\text{min}}(X_2|E) - |output| \\
\end{eqnarray*}

In this scenario, the output is revealed, and let the security parameters $\lambda_1$ and $\lambda_2$ (in bits) be the bit security to hold for keys $X_1$ and $X_2$. Given their respective distances from the uniform distribution, $\varepsilon_1$ and $\varepsilon_2$, the following conditions must be satisfied:

\begin{eqnarray}
    \label{Constraint_secuirty_param01}|output| &\leq& H^{\sqrt{2\varepsilon_1}}_{\text{min}}(X_1|E) - \lambda_1 \\
    \label{Constraint_secuirty_param02}|output| &\leq& H^{\sqrt{2\varepsilon_2}}_{\text{min}}(X_2|E) -\lambda_2
\end{eqnarray}

\subsection{Case~\ref{label4}: Reveal output AND one of the keys}\label{key_and_output_reveal}

Since the parameter $output$ is classical, the following holds (\cite{Tomamichel_2016_book}, Lemma 6.18.):

\begin{eqnarray*}
    &H^{\sqrt{2\varepsilon_1}}_{\text{min}}(X_1|E,X_2,output) \geq H^{\sqrt{2\varepsilon_1}}_{\text{min}}(X_1|E, X_2) - |output|\\
    &H^{\sqrt{2\varepsilon_2}}_{\text{min}}(X_2|E,X_1,output) \geq H^{\sqrt{2\varepsilon_2}}_{\text{min}}(X_2|E, X_1) - |output|\\
\end{eqnarray*}

And since the sources $X_1$ and $X_2$ are assumed to be independent given $E$:

\begin{eqnarray*}
    &H^{\sqrt{2\varepsilon_1}}_{\text{min}}(X_1|E,X_2,output) \geq H^{\sqrt{2\varepsilon_1}}_{\text{min}}(X_1|E) - |output|\\
    &H^{\sqrt{2\varepsilon_2}}_{\text{min}}(X_2|E,X_1,output) \geq H^{\sqrt{2\varepsilon_2}}_{\text{min}}(X_2|E) - |output|\\
\end{eqnarray*}

which is equivalent to the ``only output reveal", case~\ref{label3}.

\subsection{QKD and PQC as secure sources}\label{subsection_PQC_as_source}

Practically, source $X_1$ and/or source $X_2$ can be QKD keys, which are proven to be unconditionally secure, when no loophole in their implementation is exploited.

PQC keys could be considered to be secure sources as well, under the following explanation. Keys obtained through PQC means are assumed to have a computational smooth min-entropy conditioned to bounded adversaries $E'$ that have access to quantum computers, also known as pseudoentropy, and commonly referred to as the HILL entropy~\cite{HILL_entropy_99, Computational_Analogues_Entropy}. Another definition, dating from 1982, is given by the Yao entropy~\cite{Yao_compression_82}\cite{Computational_Analogues_Entropy}. The quantity of interest in this case is the HILL entropy, which states that if a PQC key $K_{PQC}$ of size $S_1$ is $\varepsilon_{PQC}$-indistinguishable from a true distribution with information-theoretic min-entropy of size at least $S_2 \leq S_1$ to a polynomial bounded adversary $E'$, then the output has $\varepsilon_{PQC}$-HILL pseudoentropy of at least $S_2$~\cite{Computational_Analogues_Entropy}: 

\begin{equation}\label{HILL_entropy_def}
    H^{HILL}_{\varepsilon_{PQC}}(K_{PQC}|E')\geq S_2
\end{equation}

The PQC key exchange primitive proofs always aim to prove that the final shared secrets are indistinguishable from random given a polynomial bounded adversary and the hardness of a computational problem. 

If rightly justified, the HILL entropy might be used to replace the min-entropies inside the QLHL (Theorem~\ref{LHL_lemma_mine}). Nonetheless, a more conservative approach is to use a smooth min-entropy, similarly to the secure sources approach defined in Theorem~\ref{theorem_secure_source}. for the remainder of the text this computationally secure smooth min-entropy notion is referred to as smooth HILL entropy. That is, if a PQC key $K_{PQC}$ of bit-size $S_1$ is assumed to be $\varepsilon_{PQC}$-indistinguishable from uniform random, conditioned on quantum side information $E$, then the trace distance from uniform random is upper-bounded by $\varepsilon_{PQC}$, yielding:

\begin{equation*}
    H_{\min}^{\sqrt{2\varepsilon_{PQC}}} (K_{PQC}|E)_{t=0} = S_1
\end{equation*}

where $t$ represents a time variable indicating the elapsed time since the creation of the PQC key. As $t$ increases, a quantum-bounded adversary $E$ uses the public information exchanged during key establishment to learn information about the final key $K_{PQC}$. If the public information used to establish $K_{PQC}$ is assumed to leak no more information about $K_{PQC}$ than the distinguishability parameter $\varepsilon_{PQC}$, then $t$ effectively remains zero (or $f(S_1, t) = S_1$) until any useful information, such as ciphertexts encoded with that key, is revealed by the honest parties. This construction of the smooth min-entropy for PQC keys implies that:

\begin{equation*}
    H_{\min}^{\sqrt{2\varepsilon_{PQC}}} (K_{PQC}|E)_{t} = f(S_1, t) \leq f(S_1, t=0) = S_1
\end{equation*}

For any $t \geq 0$, where $f$ is a function characterizing how the min-entropy of a PQC key $K_{PQC}$ of bit-size $S_1$ evolves over time $t$. This characterization is non-trivial because it involves units of time: it requires knowing the optimal attack a quantum-bounded adversary can perform to learn information about the PQC key, as well as the rate at which that attack can be executed.

Ideally, the PQC keys hold their smooth min-entropy close to maximal given a time $t_0$, that is:

\begin{equation*}
    H_{\min}^{\sqrt{2\varepsilon_{PQC}}} (K_{PQC}|E)_{t\leq t_0} = f(S_1, t_0) = f(S_1, t=0) - \delta = S_1 - \delta \approx S_1
\end{equation*}

Where $\delta$ is a positive real number such that $0 \leq \delta \ll 1$. Once $t > t_0$, the  smooth min-entropy of the PQC does not hold anymore and care has to be taken with the stored PQC keys. This same reasoning applies if PQC keys correspond to the input of the QLHL. It is important to remark that while $t \le t_0$, the PQC key can be assumed to remain $\varepsilon_{\mathrm{PQC}}$-secure; once $t > t_0$, that assumption no longer holds, as the $\varepsilon_{\mathrm{PQC}}$-security property is defined via the trace distance from a uniform distribution of the same length as the PQC key.

When inputting a PQC key to apply the QLHL, it is important to remark that the final security of the output is then no longer unconditional, but rather computational given the computational security assumptions in which PQC security is built upon. 

Note that when mixing a QKD key with a PQC key following the QLHL, if the PQC key is assumed to be revealed but not the QKD key, then the security remains unconditional. If the QKD key is assumed to be revealed but not the PQC key, then the security is rather computational.

Finally, the combined key obtained after applying the presented tool does not necessarily inherit the stronger security guarantee of the input keys -- the XOR method to combine keys also suffers from the same problem as discussed in~\cite{KEM_Combiners}. A way to ensure that the strongest guarantee is inherited is by adding an authentication mechanism which partially uses the derived secret keys. See for example the XOR-then-MAC approach~\cite{Hovelmanns_QKD_Oracle}, where a one time MAC is used to preserve the chosen-ciphertext attack (CCA) security of the combined key.


\subsection{Two PQC keys as secure sources}

As opposed to the case where QKD is one of the sources, no ITS can be achievable since all sources are computational secure rather than unconditionally secure. The PQC keys shall have their corresponding conditionally independent smooth HILL entropies to be fed into the QLHL (theorem~\ref{LHL_lemma_mine}).

The combination of keys from heterogeneous PQC algorithms provides a robust cryptographic combiner property. If one PQC key is compromised or revealed, the security of the combined result remains computationally bounded by the hardness assumptions of the remaining secure key. This approach provides a critical safety layer against future cryptanalytic breakthroughs; for instance, if a specific mathematical problem (e.g., lattice-based) is invalidated by a new quantum algorithm, the combined key remains secure so long as the secondary assumption (e.g., code-based) holds. Emphasis should be given on the fact that an authentication mechanism is required to ensure that the final combined key inherits the stronger security guarantee given the different key inputs~\cite{Hovelmanns_QKD_Oracle}; this construction allows to preserve the CCA security of a key input obtained through a CCA-secure KEM regardless of the other keys input which could be obtained through a chosen plaintext attack (CPA)-secure KEM, making it a good tool to combine cryptographic primitives without additional computational assumptions -- in the same fashion than the XOR method does as presented in the XOR-then-MAC approach~\cite{Hovelmanns_QKD_Oracle}.

Note that classical keys, shown to be solved by quantum algorithms in polynomial time, can also be used as secure sources if a non-quantum version of the smooth HILL entropy is associated, or if an upper bound of the classical trace distance characterizes the security of the classical key. Additionally, the presented analysis of this section can be expanded to combine more than two keys, for example to combine a classical, a PQC and a QKD key, by partitioning equally all keys to form the seed and input - all keys but one are allowed to be revealed.

The next section discusses the case where a random public seed is shared between honest users to perform the combination of secure sources. The analysis of the combination of more than two keys is rather straightforward compared to the private seed analysis presented in this section. An increase in secret output is expected since the seed is assumed to be $\varepsilon_{seed}-$secure, and is not allowed to be revealed before the generation of all the keys to be combined (case~\ref{label1} is not allowed).

\section{Strong seeded extractor with a $\varepsilon_{s}$-secure public seed: combining keys}\label{section_public_seed_mix}

A different approach that allows to combine the $\varepsilon_1$- and $\varepsilon_2$-secure keys, $X_1$ and $X_2$ respectively, is to apply the strong seeded extractor to the concatenated bitstring $X_1 \circ X_2$ with a $\varepsilon_{s}$-secure bitstring as the seed, which requires some assumptions of free choice, or free randomness~\cite{Colbeck2012}, if not generated by an unconditionally secure random number generator. The sources $X_1$ and $X_2$ are assumed to be independently conditioned to their corresponding quantum side information $E_1$ and $E_2$. The seed requires transmission through an authenticated classical channel - as typically done to apply Privacy Amplification, notably for QKD.

To prevent active interference, the public seed must be transmitted via an authenticated channel. Care must be taken to maintain mathematical independence between the randomness extractor's input and the seed. A way to ensure independence between the input and seed is to generate and send the seed by an appropriate certifiable QRNG after all sources have generated their respective keys (technically its generation can happen at any point in time and security holds if the seed remains a secret before utilization). In a hybrid QKD-PQC scenario, if the PQC key is already established while the QKD protocol is still in progress, the mixing seed can be sent simultaneously with the standard QKD privacy amplification seed; or the PQC key could be included within the QKD privacy amplification if preferred. Sending the seed before the PQC key is finalized is risky, as an adversary could attempt to influence the PQC generation to create correlations between the seed and the resulting key.

\begin{theorem}\label{theorem_QLHL_public_seed}
    Let $S$ be a $\varepsilon_S$-secure seed conditioned on quantum side information $E_S$ and $X$ be the concatenation of $\varepsilon_1$- and $\varepsilon_2$-secure keys $X_1$ and $X_2$, conditioned to quantum side information $E_1$ and $E_2$, respectively. Let $S$, $X_1$ and $X_2$ be independent from any other quantum side information. Let $X$ be the concatenation of $X_1$ and $X_2$. Finally, let $L$ be an entropy budget. Then the length of the output of a 1-almost universal$_2$ hash function-based randomness extractor is determined by the QLHL such that:
\end{theorem}

\begin{equation}
    |output| \leq |X| - L -2\log\left(\frac{1}{\varepsilon'}\right) + 2
\end{equation}

Where the final output is $(\varepsilon_S + \varepsilon_1 + \varepsilon_2 + \varepsilon')$-secure.

The $\varepsilon_S$-security implies that the seed is generated uniformly at random up to that security parameter -— that is, the seed is $\varepsilon_S$-close to uniform. The seed eventually becomes public: what matters in this case is the seed's uniformity rather than its privacy.

The entropy budget $L$ represents the maximum number of classical bits that may be revealed about the secret input $X$ (during or after its generation) without degrading the input security parameters $\varepsilon_1$ and $\varepsilon_2$. Under this assumption, as long as the output length $|output|$ is shortened accordingly to compensate for $L$, the final key achieves the full claimed security level of $(\varepsilon_S + \varepsilon_1 + \varepsilon_2 + \varepsilon')$.

\begin{proof}

In this scenario, the seed, denoted $F$, is assumed to be $\varepsilon_{S}$-close to uniform random, and its generation is independent of other any quantum side information, thus (definition~\ref{distance_from_uniform}):

\begin{equation*}
    D_u(F|E_F)_{\rho} = \min_{\sigma_{E_F} \in \mathcal{P}(\mathcal{H}_{E_F})\,:\,\text{Tr}\sigma_{E_F} = \text{Tr}\rho_{E_F}} \frac{1}{2} \left\| \rho_{FE_F} - \frac{\mathbb{I}_F}{\dim \mathcal{H}_{F}} \otimes \sigma_{E_F} \right\|_1 \leq \varepsilon_S
\end{equation*}

where $\rho_{FE_F} \in S_{\leq}(\mathcal{H}_{FE_F})$.

Additionally, the concatenation of keys, $X = X_1 \circ X_2$, is $(\varepsilon_1 + \varepsilon_2)$-secure given that the generation of $X_1$ and $X_2$ are independent from quantum side information $E_1$ and $E_2$ respectively (theorem~\ref{theorem_concatenation}), thus:

\begin{equation*}
    D_u(X|E_X)_{\rho} = \min_{\sigma_{E_X} \in \mathcal{P}(\mathcal{H}_{E_X})\,:\,\text{Tr}\sigma_{E_X} = \text{Tr}\rho_{E_X}} \frac{1}{2} \left\| \rho_{XE_X} - \frac{\mathbb{I}_X}{\dim \mathcal{H}_X} \otimes \sigma_{E_X} \right\|_1 \leq \varepsilon_1 + \varepsilon_2
\end{equation*}

Where $\rho_{XE} \in S_{\leq}(\mathcal{H}_{XE})$, $\rho_{XE} = \rho_{XE_1E_2} = \rho_{X_1E_1}\otimes\rho_{X_2E_2}$ and $\sigma_E = \sigma_{E_1} \otimes \sigma_{E_2}$ with $\text{tr}(\sigma_{E_1}) = \text{tr}(\rho_{E_1})$ and $\text{tr}(\sigma_{E_2}) = \text{tr}(\rho_{E_2})$.

First, let us derive the security of the output with no entropy budget. The density matrices of the input $\rho_{XE_X}$ and the seed $\rho_{FE_F}$ are the following classical-quantum states:

\begin{eqnarray*}
    \rho_{XE_X} = \sum_{x} \ket{x}\bra{x}_X \otimes \rho_{E_X}^{[x]} \\
    \rho_{FE_F} = \sum_{f} \ket{f}\bra{f}_F \otimes \rho_{E_F}^{[f]}
\end{eqnarray*}

Where the probability of $x$ and $f$ occurring is the trace of the
sub-normalized state $\rho_{E_X}^{[x]}$ and $\rho_{E_F}^{[f]}$, respectively, with $\rho_{E_X} = \sum_x \rho_{E_X}^{[x]}$ and $\rho_{E_F} = \sum_f \rho_{E_F}^{[f]}$.

The output density matrix $\rho_{ZFE_FE_X} \in \mathcal{S}_{\leq}(\mathcal{H}_{ZFE_FE_X})$ is also a classical-quantum state:

\begin{eqnarray*}
    \rho_{ZFE_FE_X} = \sum_{z}\sum_f \ket{z}\bra{z}_Z \otimes \ket{f}\bra{f}_F \otimes \rho_{E_F}^{[f]} \otimes   \rho_{E_X}^{[f,z]} 
\end{eqnarray*}

Where the bit size of any element $z \in Z$ is the output of the randomness extraction, as $z = f(x)$, where $f(x)$ represents the randomness extraction operation given the output $x$ and the 1-almost universal$_2$ hash function chosen by the seed $f$, and $\rho_{E_X}^{[f,z]} = \sum_{x, f(x) = z}\rho_{E_X}^{[x]}$.

To go from the input state $\rho_{FE_F} \otimes \rho_{XE_X}$ to the output state $\rho_{ZFE_FE_X}$, a completely positive trace preserving (CPTP) map is applied. The CPTP map acts on the output, seed and input registers, $Z$,  $F$ and $X$ respectively, and is denoted $(\mathcal{A}_{ZFX}\otimes \mathbb{I}_{E_FE_X})$. The matrix $\mathcal{A}_{ZFX}$ can be built using the Kraus representation where:

\begin{eqnarray*}
    (\mathcal{A}_{ZFX}\otimes \mathbb{I}_{E_FE_X})(\rho_{FE_F} \otimes \rho_{XE_X}) &=& \sum_{x'}M_{x'}(\rho_{FE_F} \otimes \rho_{XE_X})M_{x'}^\dagger\\
    &=& \rho_{ZFE_FE_X}
\end{eqnarray*}

where the Kraus operator $M_{x'} \in \mathcal{P}$ acts in the Hilbert space spanning all the possible registers of this problem, $\mathcal{H}_{FE_FZXE_X}$. An appropriate Kraus operator for this case is:

\begin{eqnarray*}
    M_{x'} = \sum_{f'} \ket{f'(x')}_Z \otimes\ket{f'}\bra{f'}_F \otimes  \bra{x'}_X \otimes \mathbb{I}_{E_F} \otimes \mathbb{I}_{E_X}
\end{eqnarray*}

To prove the security of the output $\rho_{ZFE_FE_X}$, its trace distance from uniform has to be upper-bounded by a negligible constant. For instance, the following requires to be upper-bounded:

\begin{eqnarray*}
    \min_{\sigma_{FE_FE_X} \in \mathcal{P}(\mathcal{H}_{FE_FE_X})} \frac{1}{2} \left\| \rho_{ZFE_FE_X} - \frac{\mathbb{I}_Z}{\dim \mathcal{H}_Z} \otimes \sigma_{FE_FE_X} \right\|_1
\end{eqnarray*}

Where $\text{Tr}(\sigma_{FE_FE_X}) = \text{Tr}(\rho_{FE_FE_X})$. By using the triangle inequality, the following bound is found for all possible operators $\sigma_{FE_FE_X}$:

\begin{eqnarray*}
    \left\| \rho_{ZFE_FE_X} - \frac{\mathbb{I}_Z}{\dim \mathcal{H}_Z} \otimes \sigma_{FE_FE_X} \right\|_1 \leq \left\| \rho_{ZFE_FE_X} - \tau_{ZFE_FE_X} \right\|_1 + \left\| \tau_{ZFE_FE_X} - \frac{\mathbb{I}_Z}{\dim \mathcal{H}_Z} \otimes \sigma_{FE_FE_X} \right\|_1
\end{eqnarray*}

Where $\tau_{ZFE_FE_X} \in \mathcal{S}_{\leq}(\mathcal{H}_{ZFE_FE_X})$. The expression of $\tau_{ZFE_FE_X}$ is defined as:

\begin{eqnarray*}
    \tau_{ZFE_FE_X} &=& \left(\sum_x\sum_f \ket{f(x)}\bra{f(x)}_Z \otimes \ket{f}\bra{f}_F \right) \otimes \frac{\tau_{E_F}}{\dim \mathcal{H}_F} \otimes \frac{\tau_{E_X}}{\dim \mathcal{H}_X} \\
    &=& \tau_{ZF}\otimes\frac{\tau_{E_F}}{\dim \mathcal{H}_F} \otimes \frac{\tau_{E_X}}{\dim \mathcal{H}_X}
\end{eqnarray*}

Where $\tau_{E_F} \in \mathcal{P}(\mathcal{H}_{E_F})$ and $\tau_{E_X} \in \mathcal{P}(\mathcal{H}_{E_X})$ are chosen such that $\text{Tr}(\tau_{E_F}) = \text{Tr}(\rho_{E_F})$ and $\text{Tr}(\tau_{E_X}) = \text{Tr}(\rho_{E_X})$.
By noting that:

\begin{eqnarray*}
    (\mathcal{A}_{ZFX}\otimes \mathbb{I}_{E_FE_X})(\frac{\mathbb{I}_F}{\dim \mathcal{H}_F} \otimes \tau_{E_F} \otimes \frac{\mathbb{I}_X}{\dim \mathcal{H}_X} \otimes \tau_{E_X}) = \tau_{ZFE_FE_X}
\end{eqnarray*}

The following bound holds by the contractivity of the trace distance:

\begin{eqnarray*}
    \left\| \rho_{ZFE_FE_X} - \tau_{ZFE_FE_X} \right\|_1\leq \left\| \rho_{FE_F}\otimes\rho_{XE_X} - \frac{\mathbb{I}_F}{\dim \mathcal{H}_F} \otimes \tau_{E_F} \otimes \frac{\mathbb{I}_X}{\dim \mathcal{H}_X} \otimes \tau_{E_X} \right\|_1 
\end{eqnarray*}

The right side of the inequality can be further upper bounded using the triangle inequality:

\begin{eqnarray*}
    &\left\| \rho_{FE_F}\otimes\rho_{XE_X} - \frac{\mathbb{I}_F}{\dim \mathcal{H}_F} \otimes \tau_{E_F} \otimes \frac{\mathbb{I}_X}{\dim \mathcal{H}_X} \otimes \tau_{E_X} \right\|_1\leq\\
    &\left\| \rho_{FE_F}\otimes\rho_{XE_X} - \frac{\mathbb{I}_F}{\dim \mathcal{H}_F} \otimes \tau_{E_F} \otimes \rho_{XE_X} \right\|_1 + \left\| \frac{\mathbb{I}_F}{\dim \mathcal{H}_F} \otimes \tau_{E_F} \otimes \rho_{XE_X} - \frac{\mathbb{I}_F}{\dim \mathcal{H}_F} \otimes \tau_{E_F} \otimes \frac{\mathbb{I}_X}{\dim \mathcal{H}_X} \otimes \tau_{E_X} \right\|_1 \\
    &= \left\| \rho_{FE_F} - \frac{\mathbb{I}_F}{\dim \mathcal{H}_F} \otimes \tau_{E_F} \right\|_1 + \left\| \rho_{XE_X} - \frac{\mathbb{I}_X}{\dim \mathcal{H}_X} \otimes \tau_{E_X} \right\|_1 
\end{eqnarray*}

Let $\tau_{E_F}^*$ and $\tau_{E_X}^*$ be appropriately chosen states that minimize $\left\| \rho_{FE_F} - \frac{\mathbb{I}_F}{\dim \mathcal{H}_F} \otimes \tau_{E_F} \right\|_1$ and $\left\| \rho_{XE_X} - \frac{\mathbb{I}_X}{\dim \mathcal{H}_X} \otimes \tau_{E_X} \right\|_1$, respectively. Then:

\begin{eqnarray*}
    \frac{1}{2}\left\| \rho_{ZFE_FE_X} - \tau_{ZFE_FE_X} \right\|_1\leq \varepsilon_S + \varepsilon_1 + \varepsilon_2
\end{eqnarray*}

where $\tau_{ZFE_FE_X} = \tau_{ZF}\otimes \frac{\tau_{E_F}^*}{\dim \mathcal{H}_F} \otimes \frac{\tau_{E_X}^*}{\dim \mathcal{H}_X}$.

Additionally, the term $\left\| \tau_{ZFE_FE_X} - \frac{\mathbb{I}_Z}{\dim \mathcal{H}_Z} \otimes \sigma_{FE_FE_X} \right\|_1$ can be upper bounded by using the QLHL~\cite{General_LHL_Quantum_Tomamichel_2011}. First let:

\begin{equation*}
    \sigma_{FE_FE_X} = \frac{\tau_{E_F}^*}{\text{Tr}(\tau_{E_F}^*)} \otimes \sigma_{FE_X}
\end{equation*}

where $\text{Tr}(\sigma_{FE_FE_X})$ is imposed to be equal to $\text{Tr}(\rho_{FE_FE_X})$. If $\sigma_{FE_FE_X} = \tau_{E_F}^* \otimes \sigma_{FE_X}$, then $\text{Tr}(\sigma_{FE_FE_X}) = \text{Tr}(\frac{\tau_{E_F}^*}{\text{Tr}(\tau_{E_F}^*)}) \otimes \text{Tr}(\sigma_{FE_X}) = \text{Tr}(\sigma_{FE_X})$. To maintain the equality $\text{Tr}(\sigma_{FE_FE_X}) = \text{Tr}(\rho_{FE_FE_X})$, let $\text{Tr}(\sigma_{FE_X}) =  \text{Tr}(\rho_{FE_X}) = \text{Tr}(\rho_{FE_FE_X})$. Leveraging this change, the QLHL can be applied by noting that:

\begin{eqnarray*}
    &\left\| \tau_{ZFE_FE_X} - \frac{\mathbb{I}_Z}{\dim \mathcal{H}_Z} \otimes \sigma_{FE_FE_X} \right\|_1 \\
    &= \left\| \left(\sum_x\sum_f \ket{f(x)}\bra{f(x)}_Z \otimes \ket{f}\bra{f}_F \right) \otimes \frac{\text{Tr}(\tau_{E_F}^*)\tau_{E_F}^*}{\text{Tr}(\tau_{E_F}^*)\dim \mathcal{H}_F} \otimes \frac{\tau_{E_X}^*}{\dim \mathcal{H}_X} - \frac{\mathbb{I}_Z}{\dim \mathcal{H}_Z} \otimes \frac{\tau_{E_F}^*}{\text{Tr}(\tau_{E_F}^*)} \otimes \sigma_{FE_X} \right\|_1\\
    &= \left\| \frac{\tau_{E_F}^*}{\text{Tr}(\tau_{E_F}^*)} \right\|_1 \left\| \sum_x\sum_f p_f\ket{f(x)}\bra{f(x)}_Z \otimes \ket{f}\bra{f}_F \otimes \tau_{E_X}^{*[x]} - \frac{\mathbb{I}_Z}{\dim \mathcal{H}_Z} \otimes \sigma_{FE_X} \right\|_1\\
    &= \left\| \sum_z\sum_f p_f\ket{z}\bra{z}_Z \otimes \ket{f}\bra{f}_F \otimes \tau_{E_X}^{*[f,z]} - \frac{\mathbb{I}_Z}{\dim \mathcal{H}_Z} \otimes \sigma_{FE_X} \right\|_1
\end{eqnarray*}

where $p_f = \frac{1}{\dim \mathcal{H}_F}$, $\tau_{E_X}^{*[x]} = \frac{\text{Tr}(\tau_{E_F}^*)\tau_{E_X}^*}{\dim \mathcal{H}_X}$, and $\tau_{E_X}^{*[f,z]} = \sum_{x, f(x)=z} \tau_{E_X}^{*[x]}$. With an appropriate choice of $\sigma_{FE_X} \in \mathcal{P}(\mathcal{H}_{FE_X})$, the QLHL can be applied:

\begin{eqnarray*}
    \min_{\sigma_{FE_X}}\frac{1}{2}\left\| \sum_z\sum_f p_f\ket{z}\bra{z}_Z \otimes \ket{f}\bra{f}_F \otimes \tau_{E_X}^{*[f,z]} - \frac{\mathbb{I}_Z}{\dim \mathcal{H}_Z} \otimes \sigma_{FE_X} \right\|_1 \leq \frac{1}{2}\sqrt{2^{|output| - H_{\min}(X|E_X)_{\tau}}}
\end{eqnarray*}

where $\text{Tr}(\sigma_{FE_X}) = \text{Tr}(\text{Tr}_Z(\sum_z\sum_f p_f\ket{z}\bra{z}_Z \otimes \ket{f}\bra{f}_F \otimes \tau_{E_X}^{*[f,z]}))$ is verified as proven below:

\begin{eqnarray*}
    \text{Tr}_Z(\sum_z\sum_f p_f\ket{z}\bra{z}_Z \otimes \ket{f}\bra{f}_F \otimes \tau_{E_X}^{*[f,z]}) &=& \sum_z\sum_f p_f \ket{f}\bra{f}_F \otimes \sum_{x, f(x)=z} \frac{\text{Tr}(\tau_{E_F}^*)\tau_{E_X}^*}{\dim \mathcal{H}_X} \\
    &=&\sum_f p_f \ket{f}\bra{f}_F \otimes  \frac{\text{Tr}(\tau_{E_F}^*)\tau_{E_X}^*}{\dim \mathcal{H}_X}\sum_z\sum_{x, f(x)=z} 1 \\
    &=& \sum_f p_f \ket{f}\bra{f}_F \otimes  \frac{\text{Tr}(\tau_{E_F}^*)\tau_{E_X}^*}{\dim \mathcal{H}_X} \dim\mathcal{H}_X\\
    &=& \sum_f p_f \ket{f}\bra{f}_F \otimes  \text{Tr}(\tau_{E_F}^*)\tau_{E_X}^*
\end{eqnarray*}

By tracing out the remaining of the components:

\begin{equation*}
    \text{Tr}(\sigma_{FE_X}) = \text{Tr}(\tau_{E_F}^*)\text{Tr}(\tau_{E_X}^*)
\end{equation*}

By noting that $\text{Tr}(\tau_{E_F}^*) = \text{Tr}(\rho_{E_F})$ and $\text{Tr}(\tau_{E_X}^*) = \text{Tr}(\rho_{E_X})$, it yields that $\text{Tr}(\sigma_{FE_X}) = \text{Tr}(\rho_{E_F})\text{Tr}(\rho_{E_X})$.

On the other hand, the density matrix $\sigma_{FE_X}$ is chosen such that its trace equals the trace of $\rho_{FE_FE_X}$. Note that:

\begin{eqnarray*}
    \text{Tr}(\rho_{FE_FE_X}) = \text{Tr}(\text{Tr}_Z[\text{Tr}_F(\rho_{ZFE_FE_X})])
\end{eqnarray*}

where:

\begin{eqnarray*}
    \text{Tr}_Z[\text{Tr}_F(\rho_{ZFE_FE_X})] &=& \text{Tr}_Z\left[\text{Tr}_F\left(\sum_{z}\sum_f \ket{z}\bra{z}_Z \otimes \ket{f}\bra{f}_F \otimes \rho_{E_F}^{[f]} \otimes   \rho_{E_X}^{[f,z]}\right)\right] \\
    &=& \text{Tr}_Z\left[ \sum_{z}\sum_f \ket{z}\bra{z}_Z \otimes \rho_{E_F}^{[f]} \otimes \rho_{E_X}^{[f,z]} \right]\\
    &=& \sum_{z}\sum_f \rho_{E_F}^{[f]} \otimes \rho_{E_X}^{[f,z]}
\end{eqnarray*}

And by noting that $\rho_{E_X}^{[f,z]} = \sum_{x, f(x) = z}\rho_{E_X}^{[x]}$ and that the preimages $\{x \in X \mid f(x)=z\}_{z \in \mathcal{Z}}$ form a disjoint partition of the domain $X$, summing over all elements $z \in Z$ reconstructs the sum over all inputs $x \in X$, then $\sum_z\sum_{x, f(x) = z}\rho_{E_X}^{[x]} = \sum_x \rho_{E_X}^{[x]}$, yielding:

\begin{eqnarray*}
    \text{Tr}_Z[\text{Tr}_F(\rho_{ZFE_FE_X})] &=& \left(\sum_f \rho_{E_F}^{[f]}\right) \otimes \left(\sum_x\rho_{E_X}^{[f,z]}\right)
\end{eqnarray*}

And by tracing the remainder of the register, it holds that $\text{Tr}(\sigma_{FE_X}) = \text{Tr}(\text{Tr}_Z(\sum_z\sum_f p_f\ket{z}\bra{z}_Z \otimes \ket{f}\bra{f}_F \otimes \tau_{E_X}^{*[f,z]}))$ is verified.

Moreover, the min-entropy of the input state $\tau_{XE_X}$ is maximal:

\begin{eqnarray*}
    H_{\text{min}}(X|E_X)_{\tau} &=& \max_{\sigma_E \in S_{=}(\mathcal{H}_{E_X})} \sup\left\{\lambda \in \mathbb{R}:\tau_{XE_X} \leq 2^{-\lambda}\mathbb{I}_X \otimes \sigma_{E_X}\right\}\\
    &=& \max_{\sigma_E \in S_{=}(\mathcal{H}_{E_X})} \sup\left\{\lambda \in \mathbb{R}:\frac{\mathbb{I}_X}{\dim \mathcal{H}_X} \otimes \tau_{E_X} \leq 2^{-\lambda}\mathbb{I}_X \otimes \sigma_{E_X}\right\}\\
    &=& \dim \mathcal{H}_{X} = |X|
\end{eqnarray*}

Putting everything together:

\begin{eqnarray*}
    &\min_{\sigma_{FE_FE_X \in \mathcal{P}(\mathcal{H}_{FE_FE_X})}}\left\| \rho_{ZFE_FE_X} - \frac{\mathbb{I}_Z}{\dim \mathcal{H}_Z} \otimes \sigma_{FE_FE_X} \right\|_1 \\
    &\leq \left\| \rho_{ZFE_FE_X} - \tau_{ZFE_FE_X} \right\|_1 + \left\| \tau_{ZFE_FE_X} - \frac{\mathbb{I}_Z}{\dim \mathcal{H}_Z} \otimes \sigma_{FE_FE_X} \right\|_1 \\
    &\leq \varepsilon_S + \varepsilon_1 + \varepsilon_2 + \frac{1}{2}\sqrt{2^{|output| - |X|}}
\end{eqnarray*}

Now let us take into account the leakage of $L$ bits of $X$. To do so, another CPTP map is defined, mapping $\rho_{XE_X}$ into the new input state $\rho_{XE_XE_L} \in \mathcal{S}_{\leq}(\mathcal{H}_{XE_XE_L})$ such that:

\begin{equation*}
    \rho_{XE_XE_L} = \sum_x \ket{x}\bra{x}_X \otimes \rho_{E_X}^{[x]} \otimes \ket{l(x)}\bra{l(x)}_{E_L}
\end{equation*}

where $l(x)$ represents a functions that leaks at most $L$ bits from element $x$. This construction allow us to upper-bound the new input state $\rho_{XE_XE_L}$ with the non-leakage term $\rho_{XE_X}$ as the leakage is assumed to not degrade the initial security parameter $\varepsilon_1 + \varepsilon_2$, thus, by the contractivity of the trace distance under CPTP maps:

\begin{eqnarray*}
     &\frac{1}{2} \left\| \rho_{XE_XE_L} - \sum_X\frac{\ket{x}\bra{x}}{\dim \mathcal{H}_X} \otimes \tau_{E_X}^*\otimes \ket{l(x)}\bra{l(x)}_{E_L} \right\|_1 \leq\\
     &\min_{\tau_{E_X} \in \mathcal{P}(\mathcal{H}_{E_X})} \frac{1}{2} \left\| \rho_{XE_X} - \frac{\mathbb{I}_X}{\dim \mathcal{H}_X} \otimes \tau_{E_X} \right\|_1 \leq \varepsilon_1 + \varepsilon_2    
\end{eqnarray*}

where $\sigma_{E_X}^*$ minimizes the middle term of the inequation and $\text{Tr}\sigma_{E_X} = \text{Tr}\rho_{E_X}$.

The same proof as for the non-leakage version is applied, with the following changes and with the appropriate change in the underlying Hilbert spaces:

\begin{eqnarray*}
    &\left\| \rho_{ZFE_FE_XE_L} - \frac{\mathbb{I}_Z}{\dim \mathcal{H}_Z} \otimes \sigma_{FE_FE_XqE_L} \right\|_1 \\
    &\leq \left\| \rho_{ZFE_FE_XE_L} - \tau_{ZFE_FE_XE_L} \right\|_1 + \left\| \tau_{ZFE_FE_XE_L} - \frac{\mathbb{I}_Z}{\dim \mathcal{H}_Z} \otimes \sigma_{FE_FE_XE_L} \right\|_1 \\
    &\tau_{ZFE_FE_X} \Rightarrow \tau_{ZFE_FE_XE_L}= \sum_x\sum_f \ket{f(x)}\bra{f(x)}_Z \otimes \ket{f}\bra{f}_F \otimes \frac{\tau_{E_F}}{\dim \mathcal{H}_F} \otimes \frac{\tau_{E_X} }{\dim \mathcal{H}_X} \otimes \ket{l(x)}\bra{l(x)}_{E_L} \\
    &(\mathcal{A}_{ZFXE}\otimes \mathbb{I}_{E_FE_XE_L})\left(\frac{\mathbb{I}_F}{\dim \mathcal{H}_F} \otimes \tau_{E_F} \otimes \frac{\mathbb{I}_X}{\dim \mathcal{H}_X} \otimes \tau_{E_X}\otimes \ket{l(x)}\bra{l(x)}_{E_L}\right) = \tau_{ZFE_FE_XE_L}\\
\end{eqnarray*}

where $\frac{\mathbb{I}_F}{\dim \mathcal{H}_F} \otimes \tau_{E_F} \otimes \frac{\mathbb{I}_X}{\dim \mathcal{H}_X} \otimes \tau_{E_X}\otimes \ket{l(x)}\bra{l(x)}_{E_L} = \tau_{FE_F}\otimes\tau_{XE_XE_L}$,

\begin{eqnarray*}
    &\left\| \rho_{ZFE_FE_XE_L} - \tau_{ZFE_FE_XE_L} \right\|_1 \leq \left\| \rho_{FE_FXE_XE_L} - \tau_{FE_FXE_XE_L} \right\|_1 \\
    &\leq \left\| \rho_{FE_F}\otimes\rho_{XE_XE_L} - \frac{\mathbb{I}_F}{\dim \mathcal{H}_F} \otimes \tau_{E_F}\otimes\rho_{XE_XE_L} \right\|_1 + \left\| \frac{\mathbb{I}_F}{\dim \mathcal{H}_F} \otimes \tau_{E_F} \otimes \rho_{XE_XE_L} - \tau_{FE_FXE_XE_L} \right\|_1\\
    &= \left\| \rho_{FE_F} - \frac{\mathbb{I}_F}{\dim \mathcal{H}_F} \otimes \tau_{E_F} \right\|_1 + \left\| \rho_{XE_XE_L} - \frac{\mathbb{I}_X}{\dim \mathcal{H}_X} \otimes \tau_{E_X}\otimes \ket{l(x)}\bra{l(x)}_{E_L} \right\|_1
\end{eqnarray*}

Which is upper bounded by $\varepsilon_S + \varepsilon_1 + \varepsilon_2$ with properly selected states $\tau_{E_F}^*$ and $\tau_{E_X}^*$.

Finally, the term $\left\| \tau_{ZFE_FE_XE_L} - \frac{\mathbb{I}_Z}{\dim \mathcal{H}_Z} \otimes \sigma_{FE_FE_XE_L} \right\|_1$ can be upper bounded using the QLHL once again, by using the valid transformation $\sigma_{FE_FE_XE_L} = \tau_{E_F} \otimes \sigma_{FE_XE_L}$, yielding:

\begin{eqnarray*}
    &\frac{1}{2}\left\| \tau_{ZFE_FE_XE_L} - \frac{\mathbb{I}_Z}{\dim \mathcal{H}_Z} \otimes \sigma_{FE_FE_XE_L}^* \right\|_1 \\
    &\leq\min_{\sigma_{FE_XE_L}}\frac{1}{2}\left\| \sum_z\sum_f p_f\ket{z}\bra{z}_Z \otimes \ket{f}\bra{f}_F \otimes \tau_{E_XE_L}^{*[f,z]} - \frac{\mathbb{I}_Z}{\dim \mathcal{H}_Z} \otimes \sigma_{FE_XE_L} \right\|_1 \\
    &\leq \frac{1}{2}\sqrt{2^{|output| - H_{\min}(X|E_XE_L)_{\tau}}}
\end{eqnarray*}

where $\sigma_{FE_FE_XE_L}^*$ is chosen such that the second line is minimized, where $\text{Tr}(\sigma_{FE_XE_L}) = \text{Tr}(\tau_{FE_XE_L})$, and $\tau_{E_XE_L}^{*[f,z]} = \sum_{x, f(x)=z} \tau_{E_X}^{*[x]} \otimes \ket{l(x)}\bra{l(x)}_{E_L}$.

Furthermore, given that the states from $E_L$ cannot provide more than $L$ bits of information on $X$ by definition, the min-entropy $H_{\min}(X|E_XE_L)$ is bounded by:

\begin{equation*}
    H_{\min}(X|E_XE_L) \ge H_{\min}(X|E_X) - L = |X| - L
\end{equation*}

Giving the final inequation:

\begin{eqnarray*}
    \frac{1}{2}\left\| \rho_{ZFE_FE_XE_L} - \frac{\mathbb{I}_Z}{\dim \mathcal{H}_Z} \otimes \sigma_{FE_FE_XqE_L} \right\|_1 \leq \varepsilon_S+\varepsilon_1+\varepsilon_2+\frac{1}{2}\sqrt{2^{|output| - |X| + L}}
\end{eqnarray*}

The proof reaches its end by imposing that $\frac{1}{2}\sqrt{2^{|output| - |X| + L}} \leq \varepsilon'$, for any arbitrary $\varepsilon' \ll 1$. By isolating $|output|$ from the last inequation, the theorem is found.

\end{proof}




\subsection{All cases: Adversary has control of the generation of one of the keys, or one of the keys is revealed after generation, or/and reveal of output}\label{all_cases_public}

Cases~\ref{label1} and~\ref{label2} are treated using the same argument because in the public seed scenario the reveal of the input is irrelevant as to when it happens. By using theorem~\ref{theorem_QLHL_public_seed}, the output of a 1-almost universal$_2$ hash function is $(\varepsilon_S + \varepsilon_1 + \varepsilon_2 + \varepsilon')$-secure if:

\begin{equation*}
    |output| \leq |X| - L -2\log\left(\frac{1}{\varepsilon'}\right) + 2
\end{equation*}

Since one of the keys is allowed to be fully revealed during or after generation, the following must hold:

\begin{eqnarray*}
        |output| \leq |X| - |X_1| -2\log\left(\frac{1}{\varepsilon'}\right) + 2\\
        |output| \leq |X| - |X_2| -2\log\left(\frac{1}{\varepsilon'}\right) + 2
\end{eqnarray*}

Implying that the output remains $(\varepsilon_S + \varepsilon_1 + \varepsilon_2 + \varepsilon')$-secure if:

\begin{equation}
    |output| = \left\lfloor \min(|X_1|,|X_2|) -2\log\left(\frac{1}{\varepsilon'}\right) + 2 \right\rfloor
\end{equation}








Regarding the cases~\ref{label3} and~\ref{label4}, ``reveal only the output" and ``reveal output and one of the key" scenarios respectively, the bounds are the same as in sections~\ref{only_output_reveal} and~\ref{key_and_output_reveal}. 

To finish this work, advantages over XOR and some other applications are now discussed.

\subsection{Advantages over XOR and potential application on hybrid QKD-PQC protocols}\label{Application_QKD_PQC}

The proposed key combination method introduces a trade-off: it requires a larger input key space than a standard XOR operation. By utilizing a greater volume of input material, the system establishes a larger initial min-entropy pool from which the final key is extracted. A higher compression ratio ensures that even if the entire output key is compromised, a significant portion of the source's min-entropy remains 'unspent'. This serves as a security buffer, effectively limiting the information an adversary gains regarding the underlying sources. By accounting for potential partial leakage of both the input and the output, the QLHL framework enables the quantification of the residual min-entropy in the non-revealed key material. Under sufficient compression, this remaining entropy can surpass the security bounds provided by the XOR approach.

Furthermore, the proposed method permits partial revelation of the initial key material, enabling a level of robustness that the XOR mechanism cannot achieve -- provided the compression ratio is sufficiently high. For instance, when combining a QKD key and a PQC key via XOR, any leakage from the source keys results in a direct, proportional loss of entropy in the final combined key. In contrast, our method utilizes an ``entropy budget". If the compression rate is high enough, this margin ensures that the final key remains secure, or in other words, the combined key maintains its full $\sqrt{2\epsilon}$-smooth min-entropy relative to its length, even in the presence of source material leakage, so long as that leakage remains within the designated budget.
It is important to remark that since the $\varepsilon$-ball of states is defined using the purified distance, the smoothing parameter of the min-entropy for a key that is $\varepsilon$-secure in trace distance becomes $\sqrt{2\varepsilon}$. Research in this area is fundamental to decreasing this smoothing parameter, as a larger parameter degrades the security parameters of the extracted key when applying a generalized quantum leftover hash lemma. Note that if the $\varepsilon$-balls are defined using the trace distance, then the smoothing parameter matches the original $\varepsilon$-security of the key; however, one would have to justify why this definition of the $\varepsilon$-balls is sufficient to argue security regarding the smooth min-entropy and the final output given by the QLHL. Nonetheless, a direct treatment without using smoothing parameters usually preserves the original security parameters, as done in Theorem~\ref{theorem_QLHL_public_seed}. While the public seed scenario benefits from this theorem under the right conditions, the same result could be applied to the private seed setting (notably the cases~\ref{label1} and~\ref{label2}) provided that the independence between the seed and the input is well established. This is not pursued in this work and is left for future research.

Another potential application of the proposed method to combine key material is the potential to replace some pseudo-random functions or Key Derivation Function steps in hybrid protocols, such as the ones present in the Muckle series~\cite{Muckle, MucklePlus, KEMuckle, VMuckle}, where the role of such functions is to combine different key material while binding transcripts. Indeed, the proposed method allows inclusion of the transcript within the input, with the trade-off being the need for a bigger seed, which increases accordingly with the additional length of the input.

A further application of the proposed method is its capacity to replace standard Pseudorandom Functions (PRFs) or Key Derivation Function (KDF) steps in hybrid protocols, such as those found in the Muckle series \cite{Muckle, MucklePlus, KEMuckle, VMuckle}. In these frameworks, such functions are typically employed to combine key material while ensuring transcript binding. The proposed method facilitates the direct inclusion of the protocol transcript as part of the input. While this introduces a trade-off, requiring a larger seed that scales proportionally with the additional length of the transcript, it leverages the extractor's properties to provide a unified approach to both entropy distillation and session binding. The public seed scenario presented in this section~\ref{section_public_seed_mix} is preferred and the seed generation should come from an unconditionally secure, and preferably certified, source of randomness, such as a suitable QRNG. The main advantage of this approach is to achieve unconditional security to the final key, security provided by QKD in the case the initial QKD key is assumed to not be compromised. Note that there is an increase of input key material, which can increase by as much as a factor of approximately 10, depending on how many intermediate keys are required. An outline of the replacement of the dual PRFs within the Muckle\# scheme, without adapting the security proof to the new changes, is provided in Appendix~\ref{AppendixA}, where that increase of key material is noted. Furthermore, slight modifications inspired from reference~\cite{XOR-then-MAC} are also performed, to take into account QKD IDs in the problem. Note that the message authentication codes (MACs) have to be unconditionally secure too, which can be achieved with dual universal hash functions~\cite{universal-hash-functions, light_ITS_authentication}.

\section{Conclusion}\label{conclusion}

This work provides solutions based on the Quantum Leftover Hash Lemma (QLHL) to deal with the bootstrapping of seedless Quantum Random Number Generators (QRNGs) and the resilient combination of Post-Quantum Cryptography (PQC) and Quantum Key Distribution (QKD) keys.

Given the input and seed bitstrings with conditional independence on unconditional adversary $E$, the QLHL is expanded to include the smoothing parameter of the seed and the min-entropy of the seed conditioned on $E$, giving lemma~\ref{LHL_lemma_mine}. A hash family candidate to apply the QLHL is the modified Toeplitz matrix, given its simplicity and easy implementation. The combination of the hash family and the QLHL raises a strong seeded extractor. 

QRNGs typically require an initial seed to run a postprocessing step to make the output uniform random according to some arbitrary secure parameter. Usually, this initial seed is pre-shared in order to run the first QRNG iterations and is often not mentioned on how it is generated~\cite{Postprocessing_QRNG_Toeplitz_Trevisan, QRNG_0, QRNG_1, QRNG_2}. This work gives attention to the scenario where two adjacent QRNGs have no seed to realize the post-processing step - two seedless QRNGs.  By relying on the presented strong seeded extractor, a uniform random bitstring can be extracted given the input bitstring generated by a seedless QRNG and the seed bitstring generated by the other seedless QRNG with the condition that the min-entropy of both bitstrings is high enough as discussed in section~\ref{section_one_source_extraction}. 

Strong seeded extractors serve as a robust mixing function, combining secure keys to provide provable cryptographic resilience. While a common approach is to XOR multiple keys, this method is susceptible to poor implementation or exposure risks; specifically, if the mixed output and all but one input key are compromised, the remaining key is uniquely revealed—a critical failure if that key is reused elsewhere. Alternatives like concatenation are often impractical due to size constraints, and standard Key Derivation Functions (KDFs) lack proofs for information-theoretic security~\cite{nist_combination_keys}. By contrast, employing strong seeded extractors renders the key combination process unconditionally secure. Although this requires more initial key material than XOR to produce a shorter output, it allows for the precise quantification and tuning of security parameters, as demonstrated in Sections~\ref{section_private_seed_mix} and~\ref{section_public_seed_mix}. By leveraging the QLHL and universal hash functions, the resulting security does not rely on a single point of failure. Even in scenarios where all keys but one are compromised or where traditional combining methods like XOR fail due to implementation errors, this approach maintains the integrity of the final key. Ultimately, the presented analysis provides a robust, future-proof mechanism for key management that remains secure even against an unbounded quantum adversary. A case example of combining secure keys is to use Quantum Key Distribution (QKD) and Post-Quantum Cryptography (PQC) keys. The QKD keys are provable to be secure against unbounded adversaries. However, that is not true when it comes to PQC keys. Nevertheless, by making the assumption that PQC keys have a HILL entropy~\cite{HILL_entropy_99}, the combination is still possible, where the hybrid key security also depends on computational assumptions, as discussed in section~\ref{subsection_PQC_as_source}. The presented tool is a natural candidate to extend the ITS security of the initial QKD keys to intermediate and final keys as discussed in~\ref{Application_QKD_PQC} and the example presented in Appendix~\ref{AppendixA}.

\paragraph*{Acknowledgements} JAVG thanks Vinod Nagaraja Rao, Panagiotis Papanastasiou, Marco Lucamarini and Weeraya Vichayaprasertkul for valuable discussions. JAVG has conducted this work with the support of EPSRC PhD studentship Grant EP/W524657/1. JAVG and TS have conducted this work partially with the support of ONR Grant 62909-24-1-2002.

\printbibliography

\section*{Appendix}
\appendix
\section{Outline of modified Muckle\#-like protocol}\label{AppendixA}

This appendix~\ref{AppendixA} provides an outline of a modified version of the Muckle\# protocol~\cite{KEMuckle} in Figure~\ref{fig:modified_MuckleSharp}, without entering into the details of the security proof nor the cryptographic primitives used -- although Table~\ref{tab:traffic_details} specifies what corresponds to the parameters ``$\mathsf{traffic}_i$" and ``$H_i$, for a provided integer $i$, and Table~\ref{tab:glossaryMuckleSharp} provides a chronological glossary of all the remaining parameters. The motivation behind the modifications is to provide ITS security to the session key, in the case where everything is compromised except for the QKD key, while preserving its core design which is based on TLS 1.3~\cite{rfc8446_TLS1.3}. Note that the way the QKD key is obtained involves functions taken from~\cite{XOR-then-MAC}, where the QKD key identification number $ID_{qkd}$ has to be included in the traffic to be authenticated, to avoid attacks that aim to mismatch sessions as shown in~\cite{XOR-then-MAC}. This parameter could represent one or more ID tags concatenated.

To maintain the ITS inherent in QKD-derived keys, one must avoid processing the key material through computationally-bounded functions, as this reduces the overall security guarantee down to a computational level. The proposed method enables the combination of key material from diverse sources in an unconditional and provably secure manner. By incorporating traffic into the input, the scheme is able to bind the data directly to the key material. This raises the question on how much of a compression rate and QKD key material is required initially, to ensure that all intermediate and final keys retain their ITS status.

\begin{figure}[ht]
  \centering
  \scalebox{0.55}{
  \begin{tikzpicture}\node{
  \begin{tabular}{p{8.4cm} >{\centering}p{1.5cm} p{8.4cm}}
  \centering
  Initiator & & \centering Responder \tabularnewline
  \hline
  \centering $\mathcmd{\mathsf{sk}}_I$, $cert_I$, $\mathcmd{\mathsf{SecState}}$  & & \centering $\mathcmd{\mathsf{sk}}_R$, $cert_R$, $\mathcmd{\mathsf{SecState}}$ \tabularnewline
    & & \tabularnewline
   $n_I \xleftarrow{R} \{0,1\}^{\kappa}$ & & \tabularnewline
   $\mathcmd{\mathsf{pk}}_{pq}, \mathcmd{\mathsf{sk}}_{pq} \gets \mathcmd{\mathsf{KEM}}_{pq}.{\mathsf{KGen}}(1^\kappa)$ & & \tabularnewline
    & $\xrightarrow{\makebox[2cm]{$\bm{m_1}\colon \mathcmd{\mathsf{pk}}_{pq}, n_I$}}$ & \raggedleft $n_R \xleftarrow{R} \{0,1\}^{\kappa}$ \tabularnewline
    & & \raggedleft $s_1 \xleftarrow{R} \{0,1\}^{L_1}$ \tabularnewline
    & & \raggedleft $\mathcmd{c}_{pq}, ss_{pq} \gets \mathcmd{\mathsf{KEM}}_{pq}.\mathcmd{\mathsf{Enc}}(\mathcmd{\mathsf{pk}}_{pq})$ \tabularnewline
      & &\hfill $k_{qkd}, ID_{qkd}\gets{\mathsf{GetKey}}_{qkd}(1^\kappa)$\tabularnewline
    & $\xleftarrow{\makebox[2cm]{$\bm{m_2}\colon \mathcmd{c}_{pq}, ID_{qkd}, n_R$}}$ & \tabularnewline
    $ss_{pq} \gets \mathcmd{\mathsf{KEM}}_{pq}.\mathcmd{\mathsf{Dec}}(\mathcmd{\mathsf{sk}}_{pq}, \mathcmd{c}_{pq})$ & & \tabularnewline
     $k_{qkd}\gets{\mathsf{GetKeyWithID}}_{qkd}(ID_{qkd})$ & & \tabularnewline
    \multicolumn{3}{c}{$k_{qkd_1}\|k_{qkd_2}\|k_{qkd_3}\|k_{qkd_4} \gets k_{qkd}$} \tabularnewline
    & & \tabularnewline
    \multicolumn{3}{c}{$k_{pq} \gets \ensuremath{\mathcal{F}}(ss_{pq},\mathsf{label}_1\|H_1)$} \tabularnewline
     \multicolumn{3}{c}{$k_{1} \gets \ensuremath{\mathcal{F}}(SecState,\mathsf{label}_2\|k_{pq})$} \tabularnewline
     \multicolumn{3}{c}{$k_{2} \gets \ensuremath{\mathcal{F}}(k_1,\mathsf{label}_3\|H_1)$} \tabularnewline
    & & \tabularnewline
    & & \raggedleft $IHTS\|RHTS \gets \ensuremath{\mathcal{T}_{s_1}}(k_2\|k_{qkd_1}\|\mathsf{label}_4\|\mathsf{traffic}_1)$ \tabularnewline
    & & \tabularnewline
     & $\xleftarrow{\makebox[2cm]{$\bm{m_3}\colon \{s_1, cert_R[\mathcmd{\mathsf{pk}}_R]\}_{\mathcmd{\mathsf{RHTS}}}$}}$ \tabularnewline
    $IHTS\|RHTS \gets \ensuremath{\mathcal{T}_{s_1}}(k_2\|k_{qkd_1}\|\mathsf{label}_4\|\mathsf{traffic}_1)$ & & \tabularnewline
    Verify $cert_R[\mathcmd{\mathsf{pk}}_R]$ & & \tabularnewline
 $(\mathcmd{c}_I,ss_I)\gets\mathcmd{\mathsf{KEM}}_s.\mathcmd{\mathsf{Enc}}(\mathcmd{\mathsf{pk}}_R)$ \tabularnewline
    $s_2 \xleftarrow{R} \{0,1\}^{L_2}$ & & \tabularnewline
    & $\xrightarrow{\makebox[2cm]{$\bm{m_4}\colon \{\mathcmd{c}_I\}_{\mathcmd{\mathsf{IHTS}}}$}}$ \tabularnewline
    & & \raggedleft $ss_I\gets\mathcmd{\mathsf{KEM}}_s.\mathcmd{\mathsf{Dec}}(\mathcmd{\mathsf{sk}}_R,\mathcmd{c}_I)$ \tabularnewline
     \multicolumn{3}{c}{$k_3 \gets \ensuremath{\mathcal{F}}(k_1,\mathsf{label}_5\|ss_I)$} \tabularnewline
     \multicolumn{3}{c}{$k_4 \gets \ensuremath{\mathcal{F}}(k_3,\mathsf{label}_6\|H_2)$} \tabularnewline
    & & \tabularnewline
     $IAHTS\|RAHTS \gets \ensuremath{\mathcal{T}_{s_2}}(k_4\|k_{qkd_2}\|\mathsf{label}_7\|\mathsf{traffic}_2)$&&\tabularnewline
    & & \tabularnewline
    & $\xrightarrow{\makebox[2cm]{$\bm{m_5}\colon \{s_2, cert_I[\mathcmd{\mathsf{pk}}_I]\}_{\mathcmd{\mathsf{IAHTS}}}$}}$
    &  \tabularnewline
     & & \raggedleft $IAHTS\|RAHTS \gets \ensuremath{\mathcal{T}_{s_2}}(k_4\|k_{qkd_2}\|\mathsf{label}_7\|\mathsf{traffic}_2)$\tabularnewline
    & & \raggedleft Verify $cert_I[\mathcmd{\mathsf{pk}}_I]$\tabularnewline
    & & \raggedleft $(\mathcmd{c}_R,ss_R)\gets\mathcmd{\mathsf{KEM}}_s.\mathcmd{\mathsf{Enc}}(\mathcmd{\mathsf{pk}}_I)$ \tabularnewline
    & & \raggedleft $s_3 \xleftarrow{R} \{0,1\}^{L_3}$ \tabularnewline
    & $\xleftarrow{\makebox[2cm]{$\bm{m_6}\colon \{s_3, \mathcmd{c}_R\}_{\mathcmd{\mathsf{RAHTS}}}$}}$ \tabularnewline
    $ss_R\gets\mathcmd{\mathsf{KEM}}.\mathcmd{\mathsf{Dec}}(\mathcmd{\mathsf{sk}}_I,\mathcmd{c}_R)$ \tabularnewline
     \multicolumn{3}{c}{$k_5 \gets \ensuremath{\mathcal{F}}(k_3,\mathsf{label}_8\|ss_R)$} \tabularnewline
     \multicolumn{3}{c}{$k_6 \gets \ensuremath{\mathcal{F}}(k_5,\mathsf{label}_9\|H_3)$} \tabularnewline
     \multicolumn{3}{c}{$\mathcmd{\mathsf{fk}}_I\|\mathcmd{\mathsf{fk}}_R \gets \ensuremath{\mathcal{T}_{s_3}}(k_6\|k_{qkd_3}\|\mathsf{label}_{10}\|\mathsf{traffic}_3)$} \tabularnewline
    $s_4 \xleftarrow{R} \{0,1\}^{L_4}$ & & \tabularnewline
    $\mathcmd{\mathsf{IF}} \gets\mathcmd{\mathsf{MAC}}.\mathsf{Auth}(\mathcmd{\mathsf{fk}}_I,\mathsf{traffic}_3)$ \tabularnewline
    & $\xrightarrow{\makebox[2cm]{$\bm{m_7}\colon \{s_4, \mathcmd{\mathsf{IF}}\}_{\mathcmd{\mathsf{IAHTS}}}$}}$ & \tabularnewline
    & & \raggedleft Abort if $\mathcmd{\mathsf{MAC}}.\mathsf{Ver}(\mathcmd{\mathsf{fk}}_I,\mathsf{traffic}_3,\mathcmd{\mathsf{IF}})\stackrel{?}{=}0$ \tabularnewline
    & & \raggedleft $\mathcmd{\mathsf{RF}} \gets \mathcmd{\mathsf{MAC}}.\mathsf{Auth}(\mathcmd{\mathsf{fk}}_R,\mathsf{traffic}_4)$ \tabularnewline
    & $\xleftarrow{\makebox[2cm]{$\bm{m_8}\colon \{\mathcmd{\mathsf{RF}}\}_{\mathcmd{\mathsf{RAHTS}}}$}}$ \tabularnewline
     Abort if $\mathcmd{\mathsf{MAC}}.\mathsf{Ver}(\mathcmd{\mathsf{fk}}_R,\mathsf{traffic}_4,\mathcmd{\mathsf{RF}})\stackrel{?}{=}0$ & & \tabularnewline
    & & \tabularnewline
     \multicolumn{3}{c}{$k_7 \gets \ensuremath{\mathcal{F}}(k_5,\mathsf{label}_{11}\|H_4)$} \tabularnewline
     \multicolumn{3}{c}{$IATS\|RATS\|\mathcmd{\mathsf{SecState'}} \gets \ensuremath{\mathcal{T}_{s_4}}(k_7\|k_{qkd_4}\|\mathsf{label}_{12}\|\mathsf{traffic}_5)$} \tabularnewline
  \end{tabular}
};
\end{tikzpicture}
}
\caption{Outline of modified Muckle\#-like protocol. Parameters and traffic details are provided in Tables~\ref{tab:traffic_details} and~\ref{tab:glossaryMuckleSharp}} 
\label{fig:modified_MuckleSharp}
\end{figure}



\begin{table}[ht]
    \centering
    \begin{minipage}{\textwidth}
    \centering
    \setlength{\tabcolsep}{6pt} 
    \renewcommand{\arraystretch}{1.4} 
    \footnotesize 
    \begin{tabular}{|c|c|} \hline
        \textbf{Named within the protocol} & \textbf{Actual input} \\ \hline
        $\mathsf{traffic}_1$  & $m_1\|m_2$ \\ \hline
        $\mathsf{traffic}_2$  & $m_1\|\dots\|m_4$ \\ \hline
        $\mathsf{traffic}_3$  & $m_1\|\dots\|m_6$ \\ \hline
        $\mathsf{traffic}_4$  & $m_1\|\dots\|m_7$ \\ \hline
        $\mathsf{traffic}_5$  & $m_1\|\dots\|m_8$ \\ \hline
        $H_i$  & $H(\mathsf{traffic}_i)$ \\ \hline
    \end{tabular}
    \caption{Traffic and hashed traffic details}
    \label{tab:traffic_details}
    \end{minipage}
\end{table}

\begin{table}[ht]
    \centering
    \begin{minipage}{\textwidth}
    \centering
    \setlength{\tabcolsep}{6pt} 
    \renewcommand{\arraystretch}{1.4} 
    \footnotesize 
    \begin{tabular}{|c|c|} \hline
        \textbf{Parameter} & \textbf{Description} \\ \hline
        $sk_{I/R}$  & Initiator/Responder's long-term secret key of a KEM-based PQC algorithm \\ \hline
        $cert_{I/R}$  & Certificate of the Initiator/Responder \\ \hline
        $\mathsf{SecState}$  & Secret State used to achieve post-compromise security \\ \hline
        $\kappa$  & Arbitrary security parameter \\ \hline
        $n_{I/R}$  & Initiator/Responder's nonces \\ \hline
        $\mathsf{pk}_{pq}/\mathsf{sk}_{pq}$  & Public key/Secret key of an ephemeral KEM-based PQC algorithm\\ \hline
        $s_i$  & Seed of instance $i$ of length $L_i$ to combine key material\\ \hline
        $c_{pq}/ss_{pq}$  & Ciphertext/Shared secret given by the ephemeral KEM-based PQC algorithm with $\mathsf{pk}_{pq}$\\ \hline
        $k_{qkd}/ID_{qkd}$  & QKD key/identity number requested with $\mathsf{GetKey}_{qkd}$ \\ \hline
        $k_{qkd_i}$  & Decomposition number $i$ of the initial QKD key $k_{qkd}$ \\ \hline
        $\mathsf{label}_i$  & Public and static labels for domain separation\\ \hline
        $k_{pq}$  & Expanded ephemeral PQC secret\\ \hline
        $k_1$  & Intermediate key obtained from combining $\mathsf{SecState}$ and $k_{pq}$\\ \hline
        $k_2$  & Expanded key with input $k_1$\\ \hline
        $IHTS/RHTS$  & Initiation/Responder Handshake Traffic Secret\\ \hline
        $\mathsf{pk}_I/\mathsf{pk}_R$  & Initiation/Responder's long-term public key of a KEM-based PQC algorithm\\ \hline
        $c_{I}/ss_{I}$  & Ciphertext/Shared secret given by the long-term KEM-based PQC algorithm with $\mathsf{pk}_{R}$\\ \hline
        $k_3$  & Intermediate key obtained from combining $k_1$ and $ss_I$\\ \hline
        $k_4$  & Expanded key with input $k_3$\\ \hline
        $k_{pq_{I/R}}$  & Expanded PQC secret given by the long-term PQC algorithm\\ \hline
        $IAHTS/RAHTS$  & Initiation/Responder Application Handshake Traffic Secret\\ \hline
        $c_{R}/ss_{R}$  & Ciphertext/Shared secret given by the long-term KEM-based PQC algorithm with $\mathsf{pk}_{I}$\\ \hline
        $k_5$  & Intermediate key obtained from combining $k_3$ and $ss_R$\\ \hline
        $k_6/k_7$  & Expanded key with input $k_5$\\ \hline
        $\mathsf{fk}_{I}/\mathsf{fk}_{R}$  & Initiator/Responder ITS-MAC secrets\\ \hline
        $\mathsf{IF}/\mathsf{RF}$  & Initiator/Responder ITS-MAC tags\\ \hline
        $IATS/RATS$  & Initiation/Responder Application Traffic Secret\\ \hline
        $\mathsf{SecState'}$  & New Secret State preserved for the next session\\ \hline
    \end{tabular}
    \caption{Chronological Glossary of the parameters within Figure~\ref{fig:modified_MuckleSharp}}
    \label{tab:glossaryMuckleSharp}
    \end{minipage}
\end{table}

First, let us determine the key material required to make the intermediate and final keys ITS. Let the QKD key $k_{qkd}$ be the only key material that is secure, or in other words, every PQC secret is known to an unbounded Adversary who has no access to the QKD keys. The QLHL is used 4 times to derive intermediate and final keys -- the function $\mathcal{T}_{s_i}$ is a dual universal hash function, for instance the modified Toeplitz matrix, with seed $s_i$ of size $L_i$. The seeds are assumed to be uniform random over the whole space, with the $\varepsilon_{s_1}$ security guarantee. For the first instantiation, with $i=1$, the following holds given theorem~\ref{theorem_QLHL_public_seed}:


\begin{equation*}
    |IHTS| + |RHTS| \leq  |input| - L  - 2\log\left(\frac{1}{\varepsilon'_1}\right) + 2
\end{equation*}

where $L$ represents the allowed entropy budget -- let it be zero for this example ($L=0$) as this is taken into account indirectly by directly mentioning which key is assumed to be secure. The output is $(\tilde{\varepsilon} + \varepsilon_{s_1} + \varepsilon'_1)$-secure, given the $\tilde{\varepsilon}$-security of the input, the $\varepsilon_{s_1}$-security of the seed and the arbitrary compression security level $\varepsilon'_1$. This implies that the individual keys $|IHTS|$ and $|RHTS|$ are also $(\tilde{\varepsilon} + \varepsilon_{s_1} + \varepsilon'_1)$-secure (theorem~\ref{theorem_decomposition}). Since the only key material that is secure is the QKD key $k_{qkd_1}$ (with the $\tilde{\varepsilon}$ security guarantee), then 
$|input| = |k_{qkd_1}|$. Hence:

\begin{equation*}
    |IHTS| + |RHTS| \leq  |k_{qkd_1}| - 2\left\lfloor \log\left(\frac{1}{\varepsilon'_1}\right) \right\rfloor + 2
\end{equation*}

For the instances $i=2$, $3$ and $4$ the same reasoning is applied, thus:

\begin{equation*}
    |IAHTS| + |RAHTS| \leq  |k_{qkd_2}| - 2\left\lfloor \log\left(\frac{1}{\varepsilon'_2}\right) \right\rfloor + 2
\end{equation*}
\begin{equation*}
    |\mathsf{fk}_I| + |\mathsf{fk}_R| \leq  |k_{qkd_3}| - 2\left\lfloor \log\left(\frac{1}{\varepsilon'_3}\right) \right\rfloor + 2
\end{equation*}
\begin{equation*}
    |IATS| + |RATS| + |\mathsf{SecState'}| \leq  |k_{qkd_4}| - 2\left\lfloor \log\left(\frac{1}{\varepsilon'_4}\right) \right\rfloor + 2
\end{equation*}

Thus, the minimum amount of QKD key required to make all intermediate and final keys ITS secure, in the case that QKD is not compromised, corresponds to:

\begin{equation*}
\begin{split}
    |k_{qkd}| = \Biggl\lceil |IATS| + |RATS| &+ |\mathsf{SecState'}| + |\mathsf{fk}_I| + |\mathsf{fk}_R| + |IAHTS| \\
    &+ |RAHTS| + |IHTS| + |RHTS| + 2\sum_{i=1}^4 \left(\left\lfloor \log\left(\frac{1}{\varepsilon'_i}\right) \right\rfloor - 1\right) \Biggl\rceil
\end{split}
\end{equation*}

By assuming that all intermediate and final keys have the same length $n$ and the security parameters $\varepsilon'_i$ are the same $\varepsilon'$ value for all $i$:

\begin{equation*}
    |k_{qkd}| = 9n + 8\left\lfloor \log\left(\frac{1}{\varepsilon'_i}\right) \right\rfloor - 8 
\end{equation*}

This implies that the initial QKD key material might be bigger by a factor of approximately $10$ when compared to the original Muckle\# protocol. However, the work presented here, notably in Figure~\ref{fig:modified_MuckleSharp}, claims that ITS security among all intermediate and final keys is guaranteed, assuming the QKD material is secure. 

By assuming that classical material has HILL entropy the same reasoning can be applied as when dealing with min-entropies, with the underlying assumption that the final security is rather computational and that the decrease in HILL entropy over time (or rather computational operations by any bounded adversary) is negligible given a certain range of time in which the final (and intermediate) keys need to remain secure. By assuming that PRFs or KDFs are computationally secure (function $\mathcal{F}$ in the Figure~\ref{fig:modified_MuckleSharp}), the HILL entropy of a smaller key, such as $k_1$, can be extended to a desired length to obtain bigger HILL entropy, required to make the QLHL work properly, for instance $k_2$. To preserve the computational security of the intermediate and final keys the computationally secure keys $k_2$, $k_4$, $k_6$ and $k_7$ must have a HILL entropy equal or bigger than $k_{qkd_1}$, $k_{qkd_2}$, $k_{qkd_3}$ and $k_{qkd_4}$ respectively, which provides directly the size of the output given by the PRF or KDF $\mathcal{F}$ with the corresponding computationally secure keys.

The same logic applies for the case scenario where everything but $\mathsf{SecState}$ is compromised. Note that the initial key $\mathsf{SecState}$ cannot be used to achieve ITS post-compromise security, because it is impossible to derive an ITS key $\mathsf{SecState'}$ at the end of the protocol with the same security level as the initial $\mathsf{SecState}$ key.

Knowing the size of the input of the corresponding Toeplitz matrix $\mathcal{T}$, the length of the seeds $L_1$, $L_2$, $L_3$ and $L_4$ can be calculated using equation~\ref{Toeplitz_Constr0} -- while taking into account the size of the corresponding traffic within the input. While the traffic is kept within the Toeplitz matrix inputs, note that these does not provide any security property other than domain separation as the malleability of the Toeplitz matrix can be exploited to reduce resulting output to the pure combination of the secret key material, without having as input the traffic; the traffic could be considered to be removed from the input. Compared to the Muckle\# scheme, sending the seeds through the classical channel increases the data volume to be transmitted between honest parties, which might be another undesirable change. 

It is worth mentioning that to preserve the ITS security given by the initial QKD keys, the MAC function used at the final steps of the protocol has to be ITS-secure as well, which is achievable by using a dual-universal hash function, which some of the constructions allow to have a relative big input size (order of $10^7$ to $10^8$ bits) with a small key (order of $350$ to $400$ bits respectively)~\cite{light_ITS_authentication}. Furthermore, because the intermediate keys are close to uniform random from the key material used to derive the final session keys leaking those does not compromise the security of the final session keys. Notably, when utilizing single-use ITS-MACs within the ``Send the Key in Cleartext" paradigm~\cite{delazzari2026sendkeycleartexthalving}, the requirements are further optimized: only one initialization key is required instead of two.

With the protocol presented in figure~\ref{fig:modified_MuckleSharp}, there are further case scenarios that are not explored here; for example, everything but $ss_I$ and $ss_R$ is compromised. The length of the extended PQC keys $k_{pq_I}$ and $k_{pq_R}$ can be estimated through the same reasoning used to estimate the length of $k_{qkd}$.

To conclude Appendix~\ref{AppendixA}, remarks on entity protection, the property of concealing an entity's identity (for instance certificate information), and on seed integrity checks, are provided. Typically, entity protection is achieved via Authenticated Encryption with Associated Data (AEAD), often implemented using computationally secure algorithms such as AES and HMAC-SHA. This process generally involves encrypting the plaintext certificates with AES to produce a ciphertext, which is then used to compute an HMAC tag. The concatenation of the ciphertext and the tag is then transmitted.

However, this construction lacks ITS security; an unbounded adversary without the QKD key could trivially compromise the entity protection. To achieve ITS-AEAD, one must encrypt the plaintext using a OTP and compute an ITS-MAC over the resulting ciphertext. While this requires a larger key sizes, it guarantees ITS entity protection. If Associated Data (public data) is included, the ITS-MAC must be computed over both the ciphertext and the Associated Data, to ensure integrity. Notice that in the protocol presented in Figure~\ref{fig:modified_MuckleSharp}, the seeds $s_1$, $s_2$, $s_3$ and $s_4$ are treated as Associated Data within the ITS-AEAD primitive. This ensures their integrity and prevents an adversary from swapping the random seeds for ``weak" material, such as all zeros which would cause the Toeplitz matrix to collapse and compromise the intermediate keys.

The security of this entity protection depends on the status of the QKD material and the protocol state:

\begin{itemize}
    \item With compromised QKD material and computationally bounded adversary: computationally secure entity protection is achieved only partially, mirroring the Muckle protocols. Because $\mathsf{SecState}$ is public during the initial execution, the first party to reveal their identity (certificates) is vulnerable in the first steps, as the AEAD tags are computed using only ephemeral material and the publicly known $\mathsf{SecState}$. If a protocol completes successfully then computationally secure entity protection for both sides is achieved.
    \item With secure QKD material and unbounded adversary: entity protection reaches full ITS security, as that AEAD tags are ITS in the first place, security provided by the QKD keys. Since the unbounded adversary has no access to the QKD keys, the first party revealing their entity (certificate) is not vulnerable to exposure.
\end{itemize}

An undesired change coming from the ITS entity protection feature is the increase of key material required to OTP the desired plaintext (for instance the certificates). This can be avoided by relaxing the security and encrypting the plaintext with a computationally secure algorithm such as Advances Standard Encryption (AES). However, the MAC tag has to remain ITS to preserve the ITS integrity check of the seeds, which is required to maintain the ITS property of the session keys when assuming that the QKD material has not been compromised.

Once the protocol succeeds for the first time, $\mathsf{SecState}$ is updated with secret material. This transition from a public to a secret state enables computationally secure Post-Compromise Security, ensuring that even if initial material was exposed, subsequent instances of the protocol remain secure under the assumption that the adversary is computationally bounded.

\end{document}